\documentclass[12pt]{article}
\usepackage{lscape}
\usepackage{epsfig}
\usepackage{amsmath}
\usepackage{hhline}
\usepackage{amssymb}
\usepackage{times}
\usepackage{cite}

\newlength{\dinwidth}
\newlength{\dinmargin}
\begin{document}  
\setlength{\tabcolsep}{1.5mm}

\newcommand{\pom}{{I\!\!P}}
\newcommand{\reg}{{I\!\!R}}
\newcommand{\slowpi}{\pi_{\mathit{slow}}}
\newcommand{\fiidiii}{F_2^{D(3)}}
\newcommand{\fiidiiiarg}{\fiidiii\,(\beta,\,Q^2,\,x)}
\newcommand{\n}{1.19\pm 0.06 (stat.) \pm0.07 (syst.)}
\newcommand{\nz}{1.30\pm 0.08 (stat.)^{+0.08}_{-0.14} (syst.)}
\newcommand{\fiidiiiful}{F_2^{D(4)}\,(\beta,\,Q^2,\,x,\,t)}
\newcommand{\fiipom}{\tilde F_2^D}
\newcommand{\ALPHA}{1.10\pm0.03 (stat.) \pm0.04 (syst.)}
\newcommand{\ALPHAZ}{1.15\pm0.04 (stat.)^{+0.04}_{-0.07} (syst.)}
\newcommand{\fiipomarg}{\fiipom\,(\beta,\,Q^2)}
\newcommand{\pomflux}{f_{\pom / p}}
\newcommand{\nxpom}{1.19\pm 0.06 (stat.) \pm0.07 (syst.)}
\newcommand {\gapprox}
   {\raisebox{-0.7ex}{$\stackrel {\textstyle>}{\sim}$}}
\newcommand {\lapprox}
   {\raisebox{-0.7ex}{$\stackrel {\textstyle<}{\sim}$}}
\def\gsim{\,\lower.25ex\hbox{$\scriptstyle\sim$}\kern-1.30ex%
\raise 0.55ex\hbox{$\scriptstyle >$}\,}
\def\lsim{\,\lower.25ex\hbox{$\scriptstyle\sim$}\kern-1.30ex%
\raise 0.55ex\hbox{$\scriptstyle <$}\,}
\newcommand{\pomfluxarg}{f_{\pom / p}\,(x_\pom)}
\newcommand{\dsf}{\mbox{$F_2^{D(3)}$}}
\newcommand{\dsfva}{\mbox{$F_2^{D(3)}(\beta,Q^2,x_{I\!\!P})$}}
\newcommand{\dsfvb}{\mbox{$F_2^{D(3)}(\beta,Q^2,x)$}}
\newcommand{\dsfpom}{$F_2^{I\!\!P}$}
\newcommand{\gap}{\stackrel{>}{\sim}}
\newcommand{\lap}{\stackrel{<}{\sim}}
\newcommand{\fem}{$F_2^{em}$}
\newcommand{\tsnmp}{$\tilde{\sigma}_{NC}(e^{\mp})$}
\newcommand{\tsnm}{$\tilde{\sigma}_{NC}(e^-)$}
\newcommand{\tsnp}{$\tilde{\sigma}_{NC}(e^+)$}
\newcommand{\st}{$\star$}
\newcommand{\sst}{$\star \star$}
\newcommand{\ssst}{$\star \star \star$}
\newcommand{\sssst}{$\star \star \star \star$}
\newcommand{\tw}{\theta_W}
\newcommand{\sw}{\sin{\theta_W}}
\newcommand{\cw}{\cos{\theta_W}}
\newcommand{\sww}{\sin^2{\theta_W}}
\newcommand{\cww}{\cos^2{\theta_W}}
\newcommand{\trm}{m_{\perp}}
\newcommand{\trp}{p_{\perp}}
\newcommand{\trmm}{m_{\perp}^2}
\newcommand{\trpp}{p_{\perp}^2}
\newcommand{\alp}{\alpha_s}

\newcommand{\alps}{\alpha_s}
\newcommand{\sqrts}{$\sqrt{s}$}
\newcommand{\LO}{$O(\alpha_s^0)$}
\newcommand{\Oa}{$O(\alpha_s)$}
\newcommand{\Oaa}{$O(\alpha_s^2)$}
\newcommand{\PT}{p_{\perp}}
\newcommand{\JPSI}{J/\psi}
\newcommand{\sh}{\hat{s}}
\newcommand{\uh}{\hat{u}}
\newcommand{\MP}{m_{J/\psi}}
\newcommand{\PO}{I\!\!P}
\newcommand{\xbj}{x}
\newcommand{\xpom}{x_{\PO}}
\newcommand{\zpom}{z_{\PO}}
\newcommand{\ttbs}{\char'134}
\newcommand{\xpomlo}{3\times10^{-4}}  
\newcommand{\xpomup}{0.05}  
\newcommand{\dgr}{^\circ}
\newcommand{\pbarnt}{\,\mbox{{\rm pb$^{-1}$}}}
\newcommand{\WBoson}{\mbox{$W$}}
\newcommand{\fbarn}{\,\mbox{{\rm fb}}}
\newcommand{\fbarnt}{\,\mbox{{\rm fb$^{-1}$}}}
\newcommand{\dsdx}[1]{$d\sigma\!/\!d #1\,$}
\newcommand{\eV}{\mbox{e\hspace{-0.08em}V}}
%
%
\newcommand{\qsq}{\ensuremath{Q^2} }
\newcommand{\gev}{\ensuremath{\mathrm{\;GeV}}}
\newcommand{\gevsq}{\ensuremath{\mathrm{\;GeV}^2}}
\newcommand{\et}{\ensuremath{E_t^*} }
\newcommand{\rap}{\ensuremath{\eta^*} }
\newcommand{\gp}{\ensuremath{\gamma^*}p }
\newcommand{\dsiget}{\ensuremath{{\rm d}\sigma_{ep}/{\rm d}E_t^*} }
\newcommand{\dsigrap}{\ensuremath{{\rm d}\sigma_{ep}/{\rm d}\eta^*} }

\newcommand{\dstar}{\ensuremath{D^*}}
\newcommand{\dstarp}{\ensuremath{D^{*+}}}
\newcommand{\dstarm}{\ensuremath{D^{*-}}}
\newcommand{\dstarpm}{\ensuremath{D^{*\pm}}}
\newcommand{\zDs}{\ensuremath{z(\dstar )}}
\newcommand{\Wgp}{\ensuremath{W_{\gamma p}}}
\newcommand{\ptds}{\ensuremath{p_t(\dstar )}}
\newcommand{\etads}{\ensuremath{\eta(\dstar )}}
\newcommand{\ptj}{\ensuremath{p_t(\mbox{jet})}}
\newcommand{\ptjn}[1]{\ensuremath{p_t(\mbox{jet$_{#1}$})}}
\newcommand{\etaj}{\ensuremath{\eta(\mbox{jet})}}
\newcommand{\detadsj}{\ensuremath{\eta(\dstar )\, \mbox{-}\, \etaj}}

\def\Journal#1#2#3#4{{#1} {\bf #2} (#3) #4}
\def\NCA{\em Nuovo Cimento}
\def\NIM{\em Nucl. Instrum. Methods}
\def\NIMA{{\em Nucl. Instrum. Methods} {\bf A}}
\def\NPB{{\em Nucl. Phys.}   {\bf B}}
\def\PLB{{\em Phys. Lett.}   {\bf B}}
\def\PRL{\em Phys. Rev. Lett.}
\def\PRD{{\em Phys. Rev.}    {\bf D}}
\def\ZPC{{\em Z. Phys.}      {\bf C}}
\def\EJC{{\em Eur. Phys. J.} {\bf C}}
\def\CPC{\em Comp. Phys. Commun.}

\begin{titlepage}

\noindent
\begin{flushleft}
{\tt DESY 11-084    \hfill    ISSN 0418-9833} \\
{\tt May 2011}                  \\
\end{flushleft}

\vspace{2cm}
\begin{center}
\begin{Large}

{\bf Measurement of the Diffractive Longitudinal Structure Function 
{\boldmath $F_L^D$} at HERA}

\vspace{2cm}

H1 Collaboration

\end{Large}
\end{center}

\vspace{2cm}

\begin{abstract}
  First measurements are presented of the diffractive cross section
  $\sigma_{ep \rightarrow eXY}$ at centre-of-mass energies $\sqrt{s}$ of $225$
  and $252\gev$, 
together with a
  precise new measurement at $\sqrt{s}$ of $319\gev$, using data taken
  with the H1 detector in the years $2006$ and $2007$. Together with
  previous H1 data at $\sqrt{s}$ of $301\gev$, the measurements are
  used to extract the diffractive longitudinal structure function
  $F_L^D$ in the range of photon virtualities $4.0 \leq Q^2 \leq
  44.0\gevsq$ and fractional proton longitudinal momentum loss $5
  \cdot 10^{-4} \leq \xpom \leq 3 \cdot 10^{-3}$. The measured $F_L^D$
  is compared with leading twist predictions based on diffractive
  parton densities extracted in NLO QCD fits to previous measurements
  of diffractive Deep-Inelastic Scattering and with a model which
  additionally includes a higher twist contribution derived from a
  colour dipole approach. The ratio of the diffractive cross section
  induced by longitudinally polarised photons to that for transversely
  polarised photons is extracted and compared with the analogous
  quantity for inclusive Deep-Inelastic Scattering
\end{abstract}

\vspace{1.5cm}

\begin{center}
Submitted to \EJC 
\end{center}

\end{titlepage}

\begin{flushleft}

F.D.~Aaron$^{5,48}$,           
C.~Alexa$^{5}$,                
V.~Andreev$^{25}$,             
S.~Backovic$^{30}$,            
A.~Baghdasaryan$^{38}$,        
S.~Baghdasaryan$^{38}$,        
E.~Barrelet$^{29}$,            
W.~Bartel$^{11}$,              
K.~Begzsuren$^{35}$,           
A.~Belousov$^{25}$
% ...

\bigskip{\it
% ...
$ ^{5}$ National Institute for Physics and Nuclear Engineering (NIPNE) ,
          Bucharest, Romania$^{ m}$ \\
% ...
$ ^{11}$ DESY, Hamburg, Germany \\
% ...
$ ^{25}$ Lebedev Physical Institute, Moscow, Russia$^{ e}$ \\
% ...
$ ^{29}$ LPNHE, Universit\'e Pierre et Marie Curie Paris 6,
          Universit\'e Denis Diderot Paris 7, CNRS/IN2P3, Paris, France \\
 $ ^{30}$ Faculty of Science, University of Montenegro,
          Podgorica, Montenegro$^{ n}$ \\
% ...
$ ^{35}$ Institute of Physics and Technology of the Mongolian
          Academy of Sciences, Ulaanbaatar, Mongolia \\
% ...
$ ^{38}$ Yerevan Physics Institute, Yerevan, Armenia \\
% ...
$ ^{48}$ Also at Faculty of Physics, University of Bucharest,
          Bucharest, Romania \\
% ...
$ ^e$ Supported by the Deutsche Forschungsgemeinschaft \\
% ...
$ ^m$ Supported by the Romanian National Authority for Scientific Research
      under the contract PN 09370101 \\
 $ ^n$ Partially Supported by Ministry of Science of Montenegro,
      no. 05-1/3-3352 \\
}\end{flushleft}

\newpage

\section{Introduction}

The observation that a significant subset of Deep-Inelastic Scattering
(DIS) events at HERA contain a large gap in activity in the forward
region \cite{ref1} prompted much theoretical and experimental
work. Such large rapidity gap topologies signify a colour singlet or
diffractive exchange and HERA has proved to be a rich environment for
their study. In particular, the study of diffractive DIS (DDIS), both
inclusive and exclusive, has supplied a wealth of experimental data
with a hard scale given by the photon virtuality, stimulating the
theoretical understanding of diffraction in terms of perturbative
quantum chromodynamics (QCD).

\begin{figure}[t]
\begin{center}
\includegraphics[width=0.45\columnwidth]{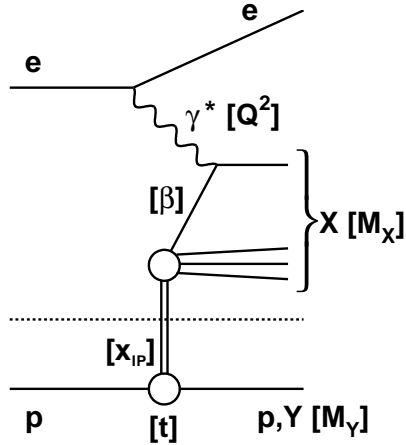}
\end{center}
\caption{A diagram of the diffractive DIS process $ep \rightarrow eXp$
  or $ep \rightarrow eXY$. The dotted line indicates where the
  diagram can be divided under the assumption of proton vertex
  factorisation.}
\label{fig:1}
\end{figure}

It has been shown that the neutral current DDIS process $ep
\rightarrow eXp$ at HERA obeys a QCD factorisation theorem
\cite{ref5}. This allows for a description of DDIS in terms of parton
densities convoluted with hard scattering matrix elements. The
diffractive parton density functions (DPDFs) depend on four kinematic
variables, so an additional assumption is often made whereby the
proton vertex dynamics factorise from the vertex of the hard scattering,
as shown in figure $\ref{fig:1}$. While this proton vertex
factorisation has no complete foundation in theory, measurements of
DDIS from both H1 \cite{ref6,ref7,ref8} and ZEUS \cite{ref9} show that
it holds well enough such that next-to-leading order (NLO)
QCD fits can be made to the data \cite{ref6,ref10,ref11,ref12}. The
DPDFs then depend only on the scale $Q^2$ and the fraction $z$ of the
total longitudinal momentum of the diffractive exchange which is
carried by the parton entering the hard scattering.

Measurements of the dijet cross section in DDIS allow tests of the
DPDFs extracted in fits to inclusive DDIS data. This process, which is
known to be dominated by boson-gluon fusion, is particularly sensitive
to the poorly known gluon DPDF at large $z$ and has 
thus been used successfully to
distinguish between different DPDF sets \cite{ref12}. DDIS events
containing charm particles in the final state have similarly been used
to test the gluon DPDF \cite{ref13}.

As in the inclusive DIS case, the cross section for DDIS can be
expressed in terms of a linear combination of structure functions,
$F_2^D$ and $F_L^D$ \cite{ref14}. While $F_2^D$ describes the total
photon-proton process, $F_L^D$ is only sensitive to the longitudinally
polarised photon contribution. As for its inclusive counterpart,
$F_L^D$ is thus zero in the quark-parton model, but may acquire a
non-zero value, $0 < F_L^D < F_2^D$ in QCD, with leading twist
contributions dependent on both the diffractive quark and gluon
densities \cite{ref15}. A measurement of $F_L^D$ provides a powerful
independent tool to verify our understanding of the underlying
dynamics of diffraction up to NLO in QCD and to test the DPDFs. This
is particularly important at the lowest $z$ values, where direct
information on the gluon density cannot be obtained from dijet data
due to kinematic limitations and where novel effects such as parton
saturation \cite{ref2} or non-DGLAP dynamics \cite{ref16,ref17} are
most likely to become important.

Previous attempts to measure $F_L^D$ \cite{ref9,ref18} have exploited
the azimuthal decorrelation between the proton and electron scattering
planes expected due to interference between the 
amplitudes for transverse and
longitudinal photon polarisations \cite{ref19}. However, due to the
relatively poor statistical precision of the measurement, the results
were consistent with zero. The H1 collaboration has recently published
measurements of the inclusive structure function $F_L(x, Q^2)$
\cite{ref20,ref21} using the centre-of-mass energy dependence of the
DIS cross section at fixed $x$ and $Q^2$. A similar approach has been
proposed to extract $F_L^D$ \cite{ref22}. 

In addition to measuring $F_L^D$ itself, it is interesting to compare
the relative sizes of the diffractive cross sections induced by
transversely and longitudinally polarised virtual photons.  This
comparison has previously been made for inclusive DIS and exclusive
vector meson production through the study of the photoabsorption
ratio, $R = \sigma_L/\sigma_T$ , where $\sigma_L$ and $\sigma_T$ are
the cross sections for the scattering of longitudinally and
transversely polarised photons, respectively. Whilst $R$ is only
weakly dependent on kinematic variables in the DIS regime for
inclusive cross sections \cite{ref20,ref23}, a strong dependence on
$Q^2$ is observed for vector meson production \cite{ref24}, the
longitudinally polarised photon cross section becoming much larger
than its transverse counterpart at large $Q^2$. Since DDIS
incorporates vector meson production and related processes at large
$z$, but exhibits kinematic dependences which are similar to those of
inclusive DIS at low $z$, it is not easy to predict its
photoabsorption ratio. By analogy with the inclusive DIS case, we
define $R^D = F_L^D /(F_2^D - F_L^D )$ for diffraction. The double
ratio $R^D/R$ thus measures the relative importance of the
longitudinally and transversely polarised photon cross sections in
diffractive compared with inclusive scattering.

In this analysis, positron-proton collision data taken at different
proton beam energies with the H1 detector at HERA in the years $2006$
and $2007$ are used to measure the diffractive 
cross section at intermediate and large inelasticities $y$. 
Dedicated low and medium energy (LME)
data with proton beam energies of $E_p = 460$ and $575\gev$ are
analysed together with data at the nominal beam energy of
$920\gev$. Previously published data at a proton beam energy $E_p =
820\gev$ \cite{ref6} are used in addition. The positron beam energy is
$27.6\gev$ in all cases. These cross sections are used to extract $F_L^D$
together with the ratio $R^D$ and the double ratio $R^D/R$.

\section{Kinematics and cross section definition}

The kinematic variables used to describe inclusive DIS are the
virtuality of the exchanged boson $Q^2$, the Bjorken scaling variable
$x$ and the inelasticity variable $y$, defined as:
\begin{equation}
Q^2 = -q^2 = -(k-k')^2
\hspace*{1cm} x=\frac{Q^2}{2P \cdot q}
\hspace*{1cm} y=\frac{P \cdot q}{P \cdot k} \ ,
\end{equation}
where $k$ and $k'$ are the four-momenta of the incoming and outgoing
positrons, respectively, and $P$ is the four-momentum of the incoming
proton. They are related to $s$, the square of the centre-of-mass
energy, by $Q^2 = sxy$.

In diffractive events, the hadronic final state can be divided into
two systems $X$ and $Y$ which are separated by the largest gap in
rapidity. A diagram for the DDIS process is shown in figure
$\ref{fig:1}$. 
The system $Y$ is either the elastically scattered proton, which is the dominant state
in the kinematic range studied here, or its low mass excitations.
In
addition to the standard DIS variables and the 
squared four-momentum transfer
at the proton vertex, $t$, the kinematic variables $\xpom$ and
$\beta$ are useful in describing the diffractive DIS interaction. They
are defined as:
\begin{equation}
\xpom = \frac{q \cdot (P-p_Y)}{q \cdot P}
\hspace*{1cm}
\beta=\frac{Q^2}{2q \cdot (P-p_Y)} \ ,
\end{equation}
where $p_Y$ is the four momentum of the elastically scattered proton
or of its low mass excitation. The variable $\xpom$ is the
longitudinal momentum fraction of the proton carried by the
diffractive exchange and $\beta$ is the longitudinal momentum fraction
of the struck quark with respect to the diffractive exchange, such
that $x = \xpom\beta$. In the simple quark-parton model, $\beta=z$,
while for higher order processes, $0 < \beta < z$. The results are
discussed in terms of a diffractive reduced cross section, $\sigma_r^D
(\beta, Q^2, \xpom)$, related to the measured differential cross
section by:
\begin{equation}
\frac{{\rm d}^3\sigma_{ep \rightarrow eXY}}{ {\rm d}\xpom {\rm d}\beta {\rm d}Q^2} = \frac{2\pi\alpha_{em}^2}{\beta Q^4} \;\cdot\; Y_+ \;\cdot\; \sigma_r^{D(3)}(\xpom, \beta, Q^2),
\label{eqn:xsec}
\end{equation}
where $Y_+ = 1 + (1-y)^2$.  The diffractive reduced cross section is
related to the diffractive structure functions by:
\begin{equation}
\sigma_r^{D(3)}(\xpom, \beta, Q^2) = F_2^{D(3)}(\xpom, \beta, Q^2) - \frac{y^2}{Y_+}F_L^{D(3)}(\xpom, \beta, Q^2).
\label{eqn:sigma}
\end{equation}
Due to the suppression term $y^2/Y_+$, the diffractive reduced cross
section is only sensitive to $F_L^D$ at large values of $y$.

As the final state system $Y$ is not measured in this analysis, the
cross section is integrated over ranges in its mass $M_Y$ and in $t$.
These ranges are chosen to be
\begin{equation}
M_Y < 1.6\gev,\; |t| < 1.0\gevsq ,
\label{eqn:myt}
\end{equation}
corresponding to the acceptance of the H1 detector in the forward direction and for
consistency with previous measurements.

\section{Models of {\boldmath {$F_L^D$}}}
\label{models}

The relationships between the diffractive structure functions and
the DPDFs have been shown to be analogous
to those of the inclusive case in the limit where the proton mass and
$t$ may be neglected compared with other relevant scales in the
interaction \cite{ref14}. The diffractive DIS structure function
$F_2^D$ is then directly sensitive to the singlet quark DPDF and the
scaling violations, $\partial F_2^D /\partial \ln Q^2$, provide a 
measure of
the gluon DPDF. NLO QCD fits to $\sigma_r^D$ at low to
intermediate $y$ values, sometimes supplemented by dijet data, thus
provide DPDFs which lead to predictions of $F_L^D$ at leading
twist. By analogy with the inclusive case \cite{ref15,ref25} and
assuming collinear factorisation \cite{ref5}, the NLO expression for
$F_L^{D(3)}$ in the $\overline{\rm MS}$ scheme is
\begin{equation}
\label{fld:nlo}
 F_L^{D(3)} (\beta,Q^2\!,\xpom\!) = \frac{\alpha_s(Q^2)}{2\pi}\! \int_\beta^1 \! dz \! \left[ \frac{4}{3} \sum_{k=\{q,\overline q \}} \!\! e_k^2 f_k\left(\frac{\beta}{z},Q^2\!,\xpom \! \right)
\! + f_g\left(\frac{\beta}{z},Q^2\!,\xpom \! \right) \! (1\!-z)\! \right],
\end{equation}
where $f_q$ and $f_g$ are the quark and gluon DPDFs and $e_k$ is the
electric charge of quark flavour $k$. At the relatively large $\beta$
values at which $F_L^D$ can be measured at HERA, both the quark and
the gluon densities are predicted to make important contributions to
$F_L^D$ , despite the dominant role played by gluons in DDIS in
general \cite{ref6,ref10}.

In this paper, the $F_L^D$ measurement is compared with predictions
derived from two NLO QCD fits to inclusive DDIS $\sigma_r^D$ data
\cite{ref6}, which are labelled `H1 2006 DPDF Fit~A' and `H1 2006 DPDF
Fit~B'. Proton vertex factorisation is assumed in both cases and the
diffractive quark densities are very similar in the two fits. However,
the two DPDF fits differ in their parameterisations of the gluon
density, which leads to considerable differences at large fractional
momenta $z$ \cite{ref6}, where the constraints from inclusive DDIS
data are poor. Corresponding differences are visible between the Fit~A
and Fit~B predictions for $F_L^D$ . The `H1 2006 DPDF Fit~B' DPDFs
give the better description of diffractive 
dijet production at HERA \cite{ref12}
and are therefore used as the default here.

A complementary approach to modelling diffractive DIS is offered by
dipole models \cite{ref3,ref4}. Viewed in the proton rest frame,
the incoming virtual photon fluctuates into a $q\bar{q}$ pair or
higher multiplicity state, whose scattering strength from the target
is governed by a universal dipole cross section. Dipole models which
are applicable to DDIS generally contain three contributions
\cite{ref4,ref26}: leading twist terms corresponding to the scattering
of $q\bar{q}$ and $q\bar{q}g$ dipoles derived from fluctuations of
transversely polarised photons, and a higher twist contribution
(suppressed like $1/Q^2$) in which $q\bar{q}$ dipoles are obtained
from longitudinally polarised photons. Dipole models thus 
tend to neglect the
leading twist contribution to $F_L^D$ which emerges naturally from NLO
DPDF fits. However, the higher twist contribution to $F_L^D$ is of
particular interest, since it can be predicted in perturbative
QCD \cite{ref27},
by coupling a $q\bar{q}$ dipole to a two-gluon exchange in a
similar phenomenology to that successfully applied to vector meson
cross sections at HERA \cite{ref28}. In many dipole-inspired models,
this higher twist component is the dominant feature of $\sigma_r^D$ at
large $\beta$ and low-to-moderate $Q^2$.

In a recent hybrid approach to fitting $\sigma_r^D$ \cite{ref29}
(labelled `Golec-Biernat \& \L uszczak' here), the leading and higher
twist contributions to $F_L^D$ are included simultaneously. A
parametrisation similar to that in \cite{ref6} is used for the
diffractive quark and gluon DPDFs, but the higher twist longitudinal
photon contribution is also included via the parametrisation employed
in \cite{ref4}. The quality of the fit to the $\sigma_r^D$ data is
similar with and without the higher twist term. However, its inclusion
leads to a sizeable effect on the diffractive gluon density at large
fractional momenta and the higher twist contribution dominates the
resulting predictions for $F_L^D$ for $\beta \gsim 0.6$ at the lowest
$Q^2$ values considered here.

\section{Experimental Method}

\subsection{H1 detector}
A detailed description of the H1 detector can be found elsewhere
\cite{ref30} and only the components essential to the present analysis
are briefly described here. The origin of the H1 coordinate system is
the nominal $ep$ interaction point at the centre of the detector, with
the direction of the proton beam defining the positive $z$-axis
(forward direction). The polar angle ($\theta$) is defined with
respect to this axis and the pseudorapidity is defined as $\eta = -\ln
\tan(\theta/2)$. The azimuthal angle $\phi$ defines the particle
direction in the transverse plane.

The analysis uses several of the tracking detectors of H1, relying
primarily on the two concentric central jet chambers (CJC) and the
central silicon tracker (CST)~\cite{cst}, which measure the transverse momenta of
charged particles in the angular range $20^{\circ} < \theta <
160^{\circ}$, together with the backward silicon tracker (BST), which
is positioned around the beam-pipe in the backward
direction. Complementary tracking information is obtained from the $z$
drift chamber COZ, which is located in between the two cylinders of
the CJC, the forward silicon tracker (FST) and the forward tracking
detector (FTD). The central inner proportional chamber (CIP)~\cite{Becker:2007ms} provides
trigger information on central tracks, the FST and BST are used to
improve the overall vertex reconstruction and the FTD is used to
improve the hadronic final state reconstruction of low momentum
particles in the forward direction.

In the backward region $-4.0 < \eta < -1.4$, a lead-scintillating
fibre calorimeter (SpaCal) is used for the identification and
measurement of the scattered positron, with an energy resolution for
electromagnetic showers of $\sigma(E)/E \simeq 7.1\%/\sqrt{E/{\rm
    GeV}} \oplus 1\%$.  Importantly, it also provides a trigger down
to positron energies of $2\gev$. The hadronic section of the SpaCal is
used in the reconstruction of the hadronic final state, especially at
the high~$y$ values accessed in this analysis. The liquid argon (LAr)
calorimeter covers the range $-1.5 < \eta < 3.4$ and is 
also used in this
analysis in the reconstruction of the hadronic
final state. It has an energy resolution 
of $\sigma(E)/E \simeq 50\%/\sqrt{E/{\rm GeV}}$ 
for hadronic showers, as obtained from
test beam measurements~\cite{ref31}.

Several of the forward detectors of H1 are used in conjunction with
the LAr to determine whether or not an event contains a large rapidity
gap close to the outgoing proton direction. The forward muon detector
(FMD) comprises two sets of three drift chambers, separated by a
toroidal magnet, covering the range $1.9 < \eta < 3.7$. Only the three
layers closest to the interaction region are considered in this
analysis. A dedicated reconstruction algorithm efficiently detects
secondary particles produced through the interactions of proton
dissociation products with the beam-pipe or other accelerator elements,
giving the FMD an effective coverage extending to around $\eta =
6.5$. The Plug is a calorimeter consisting of four double layers of
scintillator and lead absorber, read out by photomultipliers. It is
situated at $z = 4.9 \ {\rm m}$ and covers the range $3.5 < \eta < 5.5$. The
final forward detector component used in the analysis is one station
of the forward tagging system (FTS), consisting of scintillators
situated around the beam-pipe at $z = 28 \ {\rm m}$ 
covering approximately $6.0
< \eta < 7.5$.

Positrons scattered through very small polar angles can be detected
with a calorimeter (ETAG) placed at $z = -6 \ {\rm m}$ downstream in the
positron beam direction. The luminosity is determined from the
Bethe-Heitler scattering process, which is measured using a photon
calorimeter at $z = -103 \ {\rm m}$.

\subsection{Data samples }

Three samples are analysed to provide data at different centre-of-mass
energies in different kinematic ranges, as shown in table
$\ref{tab:1}$.

\begin{table}[h]
\centering
\begin{tabular}{|c|c|r@{$Q^2$}l|r@{$y$}l|r@{$.$}l|}
\hline 
\multicolumn{1}{|c|}{$E_p$} & \multicolumn{1}{|c|}{$\sqrt{s}$} & \multicolumn{2}{|c|}{$Q^2$ range} & \multicolumn{2}{|c|}{$y$ range} & \multicolumn{2}{|c|}{Luminosity}  \\
\multicolumn{1}{|c|}{$({\rm GeV})$} & \multicolumn{1}{|c|}{$({\rm GeV})$} & \multicolumn{2}{|c|}{$({\rm GeV^2})$} & \multicolumn{2}{|c|}{}  & \multicolumn{2}{|c|}{$({\rm pb^{-1}})$}  \\
\hline \hline
$460$ & $225$ & $2.5 <$ \ & \ $< 100$ & $0.1 <$ \ & \ $< 0.9$ & $8$ & $5$ \\
$575$ & $252$ & $2.5 <$ \ & \ $< 100$ & $0.1 <$ \ & \ $< 0.9$ & $5$ & $2$ \\
$920$ & $319$ & $7.0 <$ \ & \ $< 100$ & $0.1 <$ \ & \ $< 0.56$ &$126$ & $8$ \\ 
\hline
\end{tabular} 
\caption{Summary of the data samples used in the analysis.}
\label{tab:1}
\end{table}

In addition to these data, cross section measurements at $E_p = 820\gev$ from
a previous H1 publication \cite{ref6} are used to extract $F_L^D$ in
the same kinematic range.

\subsection{Event selection}

Dedicated `high~$y$' triggers are used for the LME datasets in order
to allow triggering on energy depositions as low as $2\gev$ in the
SpaCal. For $y > 0.6 \ (0.56)$ in the $460$ ($575$) ${\rm GeV}$ data, 
the SpaCal trigger decision is combined with
information from the BST or CIP in order to reduce the rate. For
lower $y$ values, corresponding to
high energy depositions in the SpaCal, triggers based on SpaCal-only
information are used for all three datasets. The combined efficiency
of the LME high~$y$ triggers is around $99\%$ for positron energies
above $3\gev$, as monitored with independent triggers. The data are
corrected for this inefficiency, which has a small dependence on the
radial position of the scattered positron in the SpaCal, $R_{spacal}$,
due to the track requirement. The combination of SpaCal-only triggers
used has a negligibly small inefficiency.

The event selection is based on the identification of the scattered
positron as a localised energy deposition, a cluster, of more than
$3.4(12.0)\gev$ in the SpaCal in the LME ($920\gev$) data. Backgrounds
due predominantly to photoproduction processes, where the scattered
positron is lost down the beam-pipe, are reduced by requiring that the
logarithmic energy-weighted cluster radius, $r_{log}$, is smaller than
$5$ cm and that the energy measured in the hadronic section of the
SpaCal associated with the cluster is less than $15\%$ of the cluster
energy. If the highest energy cluster fails to fulfill these selection
criteria, the second and third highest energy clusters
are considered in turn. QED Compton contributions, $ep
\rightarrow e\gamma p$, are suppressed by rejecting events with two
back-to-back clusters.

For the LME data, the background is further reduced by demanding a
`linked track' that can be extrapolated to the SpaCal cluster within a
radial distance of $3$ cm. The linked track is reconstructed using a
dedicated algorithm incorporating information from both the CJC and
the BST \cite{ref32}. Geometrical cuts are
applied to keep the tracking acceptance high and 
track quality requirements are applied, reflecting the 
geometry of these detectors.

In order to further reject background, a reconstructed event vertex is
required to lie within $35$ cm of the nominal interaction point for
all data samples. In order to guarantee a high vertex-finding
efficiency, the measurement is restricted to the kinematic range $y >
0.1$. An algorithm combining calorimeter and tracking information,
which optimises precision while avoiding double-counting, is used to
reconstruct the four vector of the hadronic final state (HFS)
particles \cite{ref33}. For all datasets, the quantity $\Sigma_i(E -
p_z)_i$, where the sum is over the energy $E$ minus the longitudinal
momentum $p_z$ of all final state particles including the scattered
positron, is required to be greater than $35\gev$. This quantity
should peak at twice the incident positron energy, i.e. $55\gev$, for
fully reconstructed DIS and DDIS events alike. This completes the
background rejection criteria of the inclusive event selection.

At low positron energies, the photoproduction background remains large
after all cuts. Following the procedure explained in \cite{ref20},
this residual background is estimated from 
the number of events $N_{WC}$ passing the full
analysis selection and having a negatively charged track linked to the
SpaCal cluster. The photoproduction background is expected
to be approximately charge symmetric and therefore corresponds to
approximately
$2N_{WC}$. However, a small asymmetry in its charge composition has
previously been measured \cite{ref20}. Thus the photoproduction
estimate is $1.98N_{WC}$, which is statistically subtracted from the
sample.

Diffractive DIS events are selected as a subsample of the inclusive
DIS event sample on the basis of a large rapidity gap in the forward
direction. The pseudorapidity $\eta_{max}$
of the forward-most energy deposit above
$800$ MeV in the LAr calorimeter is required to be less
than $3.3$. In addition, the FMD, Plug and FTS are required to have
no discernible signal above their typical noise levels. The combined
efficiency for rejecting proton dissociative events with $M_Y\gsim
1.6\gev$ is greater than $99\%$. These requirements 
select a subsample of events
where the hadronic final state is separated into two systems $X$ and
$Y$ by a large rapidity gap. The system $Y$, which is predominantly a
single proton, escapes undetected down the beam-pipe, while the system
$X$ is fully contained in the main H1 detector.

In order to maintain a high efficiency for the vertex reconstruction
of the DDIS event sample, an additional fiducial cut is required 
to avoid cases where both the final state system $X$ and the
positron are outside the acceptance of the CJC. The region where both
$R_{spacal} < 40 \ {\rm cm}$ and $\eta_{max} < -1.7$ is removed from the
analysis, after which the vertex-efficiency is high and well
understood throughout the measured phase space. Finally, there must be
at least one reconstructed HFS particle to define the system
$X$.

The inclusive DIS event kinematics are reconstructed using different
methods depending on the $y$ range of a given dataset. For the LME
data, only information from the reconstructed scattered positron is
used, as this method has the best resolution at large $y$:
\begin{equation}
y = 1 - \frac{E_e^{'}}{E_e} \sin^2(\frac{\theta_e}{2}) 
\hspace*{1cm} 
Q^2 = \frac{E_e^{'\;2} \sin^2(\theta_e)}{1-y} 
\hspace*{1cm}
x=\frac{Q^2}{sy} \ .
\end{equation}
Here, $E_e$ is the energy of the incident positron and $E_e^{'}$ and
$\theta_e$ are the energy and polar angle of the scattered positron,
respectively.
For the $920\gev$ data, a method with better performance at low $y$ is
used \cite{ref34}:
\begin{equation}
y = y_e^2 + y_d(1-y_d)
\hspace*{1cm} 
Q^2 = \frac{4E_e^2(1-y)}{\tan^2{\theta_e/2}} 
\hspace*{1cm}
x=\frac{Q^2}{sy} ,
\end{equation}
where $y_d=\tan{(\gamma/2)}/[\tan{(\theta_e/2)}+\tan{(\gamma/2})]$ and
$\gamma$ is the polar angle of the hadronic final state.

The four momentum of the final state system $X$ is reconstructed as
the vector sum of all HFS particles. Its mass $M_X$ is reconstructed as:
\begin{equation}
M_X= f(\eta_{max}) \sqrt {(E^2 -p_z^2 -p_x^2-p_y^2)_{_{HFS}} \frac{y}{y_h} },
\end{equation}
where $(E,p_x,p_y,p_z)_{_{HFS}}$ denotes the four vector of the HFS and
$y_h = (E-p_z)_{_{HFS}} / 2E_e$.  The term $y/y_h$ improves the resolution
and the function $f(\eta_{max})$ is determined from simulation and
corrects for detector losses.  The diffractive variables are then
reconstructed as:
\begin{equation}
\beta = \frac{Q^2}{Q^2 + M_X^2} 
\hspace*{1cm} 
\xpom = \frac{x}{\beta}.
\end{equation}

\subsection{Corrections to the data and simulations}
\label{subsec:datamc}

Monte Carlo (MC) simulations are used to correct the data for the
detector effects of acceptance, inefficiencies, and migrations between
measurement intervals. The DDIS signal is modelled for $\xpom < 0.15$
using the RAPGAP \cite{ref35} generator, with H1 2006 DPDF Fit~B
\cite{ref6} as the input DPDFs. Higher order QCD radiation is modelled
using initial and final state parton showers in the leading
$\log(Q^2)$ approximation \cite{ref37}. Hadronisation is simulated
using the Lund string model \cite{ref38} as implemented in PYTHIA
\cite{ref39}. As RAPGAP is a leading order MC generator simulating
only $F_2^D$, the effect of $F_L^D$ has been simulated by weighting
RAPGAP events by the ratio $\sigma_r^D /F_2^D$ as given at NLO by H1
2006 DPDF Fit~B. This is important at high~$y$ in order to describe
the data. At low $Q^2$, H1 2006 DPDF Fit~B undershoots the data, as
observed previously \cite{ref6}. RAPGAP is therefore reweighted
for $Q^2 < 7\gevsq$ by a parametrisation of the ratio of the previous
data to H1 2006 DPDF Fit~B. Resonant contributions to the diffractive
cross section, important at low $Q^2$ and low $M_X < 5\gev$, are
modelled using the DIFFVM \cite{ref36} generator. The DIFFVM generator
is also used to simulate proton dissociative events with $M_Y < 5\gev$
to correct the measurements to the $M_Y$ and $t$ ranges given in
equation $\ref{eqn:myt}$ under the assumption of proton vertex
factorisation. The small non-diffractive DIS background from $\xpom >
0.15$ or $M_Y > 5\gev$ is modelled using DJANGO \cite{ref40}, while
the COMPTON program \cite{ref41} is used to model the QED Compton
process, important at very low $M_X$.

The generated events are passed through a full GEANT \cite{ref42}
simulation of the H1 detector. The simulated events are subjected to
the same reconstruction and analysis chain as the data. More details
of the analysis can be found in \cite{ref43}.

\begin{figure}[htbp]
  \begin{center}
    \includegraphics[width=0.49\textwidth]{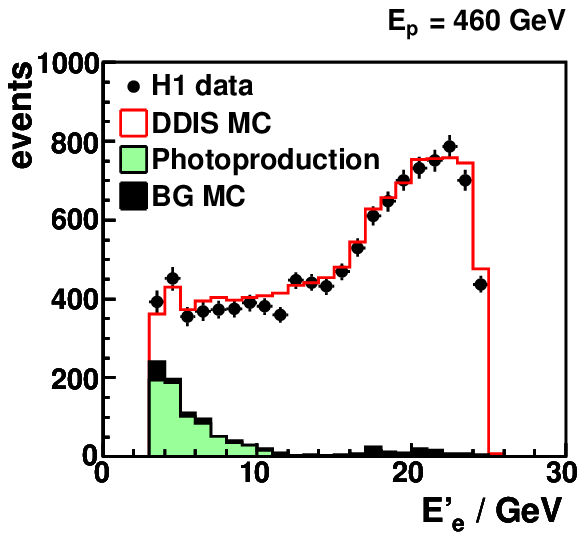}
    \includegraphics[width=0.49\textwidth]{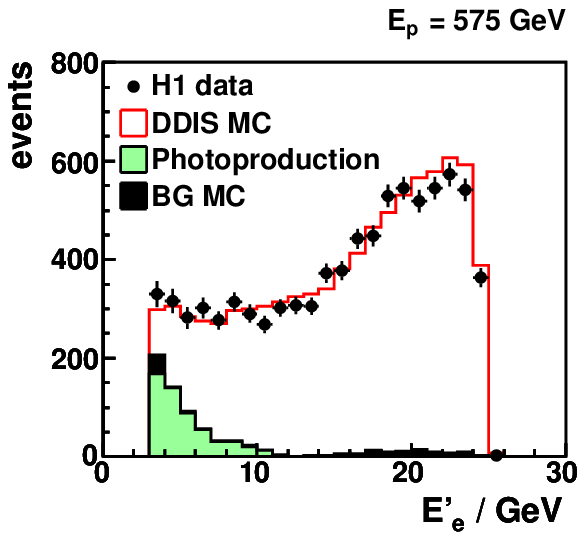}
  \end{center}
  \caption{ The energy distributions of the scattered positron 
candidates for the
    $460\gev$ (left) and $575\gev$ (right) data. The data shown as
    points are compared with the sum of the diffrective DIS MC simulation and
    background estimates (open histogram). The light-filled histogram
    shows the photoproduction background estimate from data, the
    dark-filled histogram is the sum of the QED Compton and inclusive DIS
    backgrounds, taken from MC simulations.}
\label{fig:2}
\end{figure}

Figure $\ref{fig:2}$ shows the energy distributions for positron
candidates in the LME datasets. In addition to the simulation
described above, the photoproduction estimate using the number of
candidates with the wrong charge, and the total 
background expectation are also
shown. The data are well described down to positron energies of $3.4\gev$.

\begin{figure}[htbp]
  \begin{center}
    \includegraphics[width=0.49\textwidth]{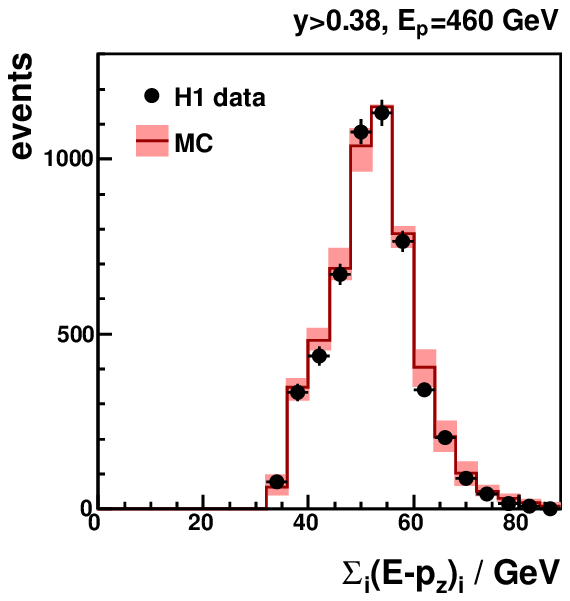}
    \includegraphics[width=0.49\textwidth]{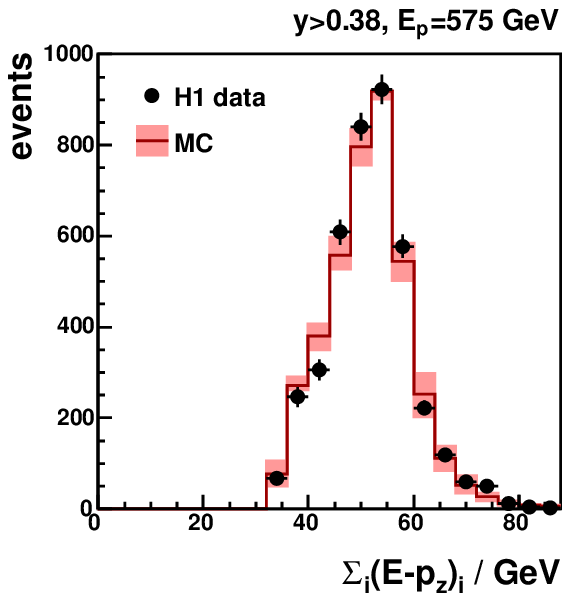}
  \end{center}
  \caption{The quantity $\Sigma_i(E-p_z)_i$ summed over all final
    state particles for the $460\gev$ (left) and $575\gev$ (right)
    data at high~$y$. The data after background subtraction are shown
    as points, compared with the MC simulation shown as a histogram.
    The shaded area shows the effect of a 
    variation of the hadronic SpaCal energy
    scale by its uncertainty of $5\%$.}
\label{fig:3}
\end{figure}

The quality of the calibration of the system $X$, in the sensitive
region at high~$y$, is illustrated in figure $\ref{fig:3}$, where
$\Sigma_i(E - p_z)_i$ peaks at the 
expected value of $55 \ {\rm GeV}$ and is well described
by the simulation.  At large $y$, the hadronic energy measurement is
strongly influenced by the hadronic energy response of the SpaCal,
which has been calibrated using inclusive DIS events \cite{ref43}. The
influence of varying the SpaCal hadronic energy scale by $\pm5\%$ is
indicated in the figure.

\begin{figure}[htbp]
  \begin{center}
    \includegraphics[width=0.32\textwidth]{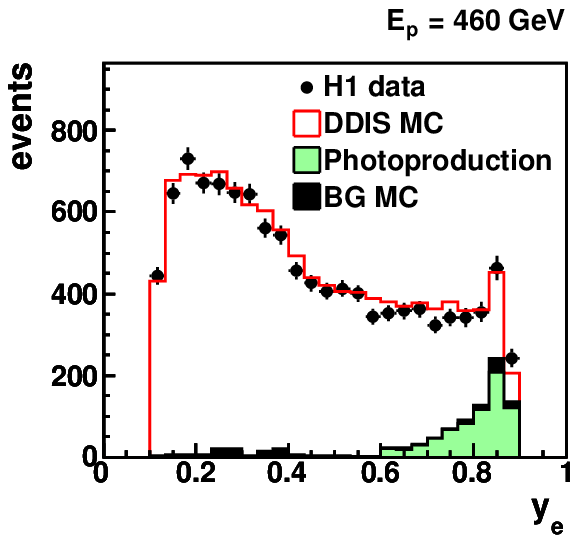}
    \includegraphics[width=0.32\textwidth]{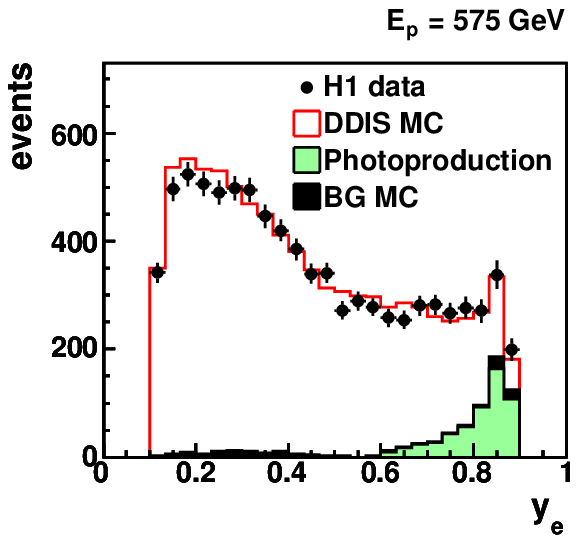}
    \includegraphics[width=0.32\textwidth]{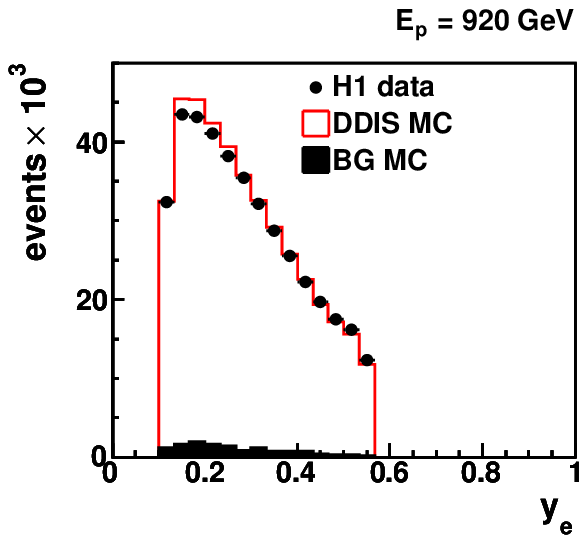}
    \includegraphics[width=0.32\textwidth]{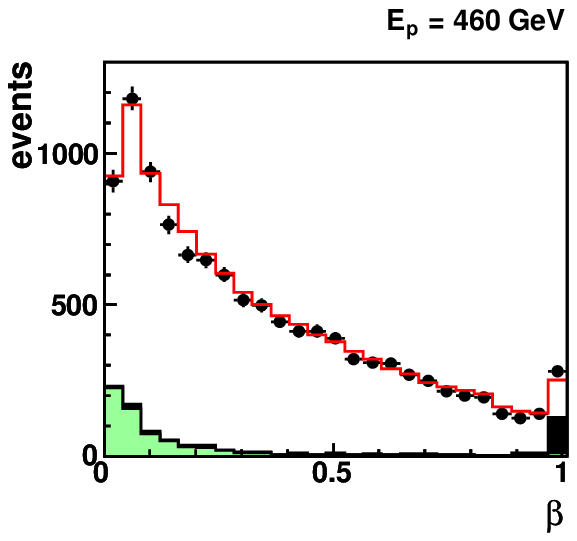}
    \includegraphics[width=0.32\textwidth]{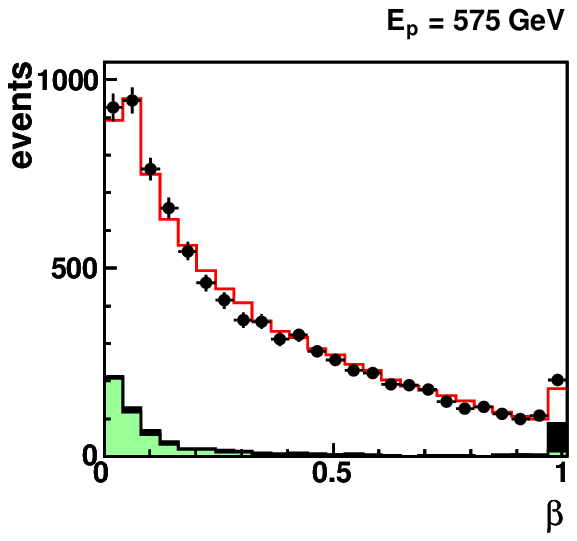}
    \includegraphics[width=0.32\textwidth]{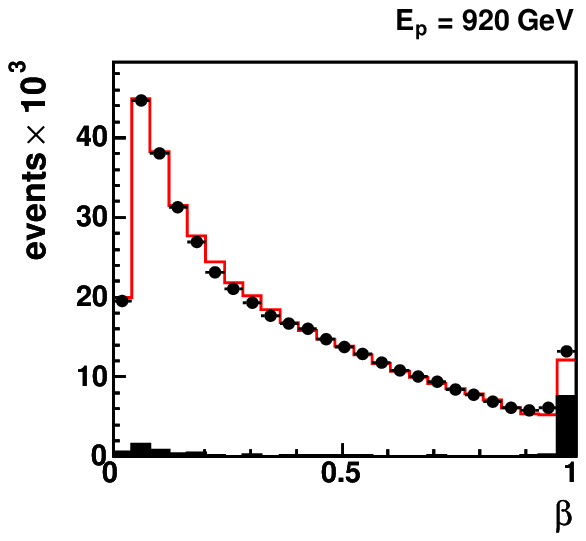}
    \includegraphics[width=0.32\textwidth]{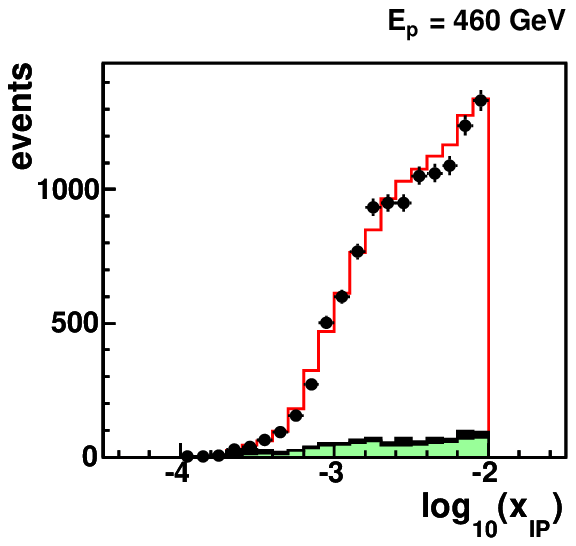}
    \includegraphics[width=0.32\textwidth]{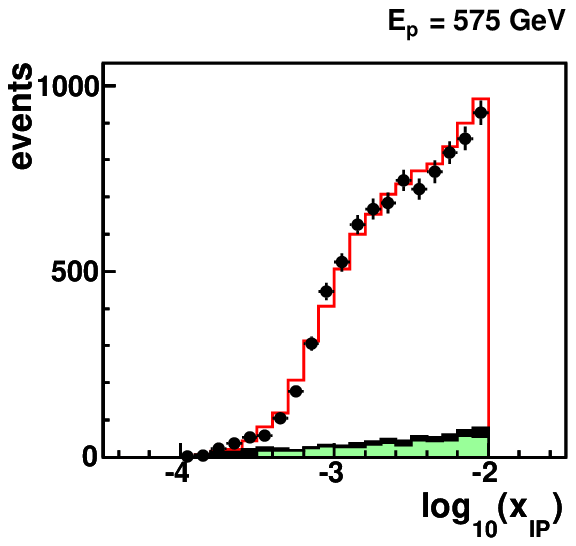}
    \includegraphics[width=0.32\textwidth]{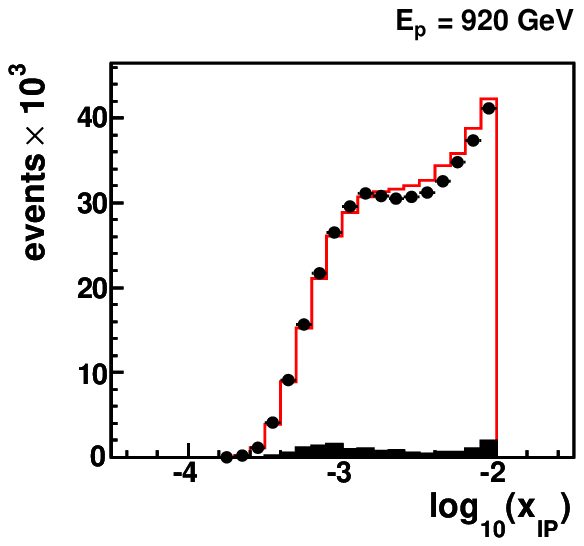}
  \end{center}
  \caption{Distributions of the kinematic quantities $y$ (top),
    $\beta$ (middle) and $\log(\xpom)$ (bottom) for the $460 \gev$
    (left), $575 \gev$ (middle) and $920 \gev$ (right) datasets.  The
    data are shown as points compared with the sum of the MC
    simulation and background estimates (open histogram).  The
    light-filled histogram shows the photoproduction background
    estimate from data, the dark-filled histogram is the sum of the QED
    Compton and inclusive DIS backgrounds, obtained from MC simulations.}
\label{fig:4}
\end{figure}

The $y, \beta$ and $\log(\xpom)$ distributions in the data are compared
with the total expectation in figure $\ref{fig:4}$ for all three
datasets. Again, the photoproduction estimate and 
the sum of all other background
sources are also shown. The quality of the description is good in
all cases.

\subsection{Cross section extraction}
\label{subsec:xsec}

The data are analysed in two $Q^2$ ranges. For $Q^2 > 7\gevsq$, data
are available from all three datasets at $E_p = 460, 575$ and
$920\gev$. For $2.5 < Q^2 < 7.0\gevsq$, only data from the $460$ and
$575\gev$ datasets are analysed. Previous measurements at $E_p =
820\gev$ \cite{ref6} are used in addition in the $Q^2$ and $\xpom$
range of the LME data. The $Q^2$, $\xpom$ and $\beta$ values of these
published data have been adjusted to the values of the current
analysis using a parameterisation of $\sigma_r^D$
derived from H1 2006 Fit~B, a procedure which results in a
systematic uncertainty of $1\%$ at $\xpom = 0.003$ and $3\%$ at $\xpom
= 0.0005$. The reduced cross section is extracted 
as a function of $\beta$, $Q^2$ and $\xpom$
from measurements of
the differential cross section according to equation $\ref{eqn:xsec}$. 
The $Q^2$
and $\xpom$ measurement intervals are large and have been optimised
for the extraction of $F_L^D$ in as broad a kinematic range as
possible.

The data are corrected for efficiencies and migrations between
measurement intervals using the MC simulation described in section
$\ref{subsec:datamc}$. The acceptance, as calculated from the MC
model, is required to be above $20\%$ for all points and is much
larger than this except at the lowest $Q^2$ and $\xpom$. Purity and
stability\footnote{ Purity is defined as the fraction of
  reconstructed MC events in a measurement interval which also
  originated in the same interval at the hadron level. Stability is the
  fraction of MC events in a measurement interval at the hadron level
  which are also reconstructed in that interval.}  are larger than
$50\%$ in all bins. For the LME data, the estimate of the
photoproduction background using the number of candidates with the
wrong charge, $N_{WC}$, is subtracted bin-by-bin for $y > 0.6$, while
below this value the background is negligible.  Inclusive DIS and
QED-Compton contributions are also subtracted bin-by-bin using the MC
simulations described in section $\ref{subsec:datamc}$. The
parametrisation of $\sigma_r^D$ using H1 2006 DPDF Fit~B is used to
correct the data to the central $Q^2$, $\xpom$ and $\beta$ values
quoted. As $\beta \rightarrow 1$, the shape of the cross section is
largely unconstrained by data and varies quickly due to resonant
contributions, making the correction to a single point in the phase
space problematic. Thus, for $\beta > 0.9$, the average cross section
in that interval is given.

The diffractive reduced cross section is integrated over the $M_Y$ and
$t$ ranges given in equation $\ref{eqn:myt}$. DIFFVM is used to
calculate the correction to this phase space, which varies with proton
beam energy. The correction factors are $1.04$, $1.06$ and $1.15$ for
the $460$, $575$ and $920\gev$ data, respectively.

For use in forming the ratio $R^D/R$, inclusive cross sections are
measured in the same binning scheme as is used for the diffractive
measurement, using the procedure 
described in \cite{ref20}. As the statistics for
the inclusive DIS sample are larger, the background subtraction is more
sophisticated. The number $N_T$ of events passing the full analysis
selection and having a signal in the ETAG photoproduction tagger and a
negatively charged linked track associated to a SpaCal cluster
provides another estimate of the photoproduction background. For the
$460 (575)\gev$ data, at low $y < 0.6(0.56)$, the photoproduction
estimate uses $N_T$,
while
for higher $y$ the photoproduction background is estimated using the
number of candidates with the wrong charge, $N_{WC}$. For the
$920\gev$ data, the estimate based on
positron-tagged events is used for all $y$.

\subsection{Systematic uncertainties}

A full systematic error analysis is performed, which carefully
considers correlations between measurement intervals and data at
different centre-of-mass energies.
The sources of systematic uncertainty that have correlations between
cross section measurement points at different $E_p$ values are as
follows.

\begin{itemize}
\item The uncertainty on the electromagnetic energy scale of the
  SpaCal is $0.2\%$ at the kinematic peak of $E_e^{'}=27.6\gev$,
  increasing linearly such that it would be $1\%$ at $E_e^{'}=1\gev$.

\item The possible bias in $\theta_e$ is estimated using the
  mean difference in polar angle between the linked track and the SpaCal
  cluster, which is measured to be less than $1$ mrad.

\item Noise is simulated in the LAr calorimeter using
  randomly-triggered events. The fraction of energy identified and
  subtracted as noise is known to a precision of $15\%$.

\item The hadronic section of the SpaCal is calibrated to a precision
  of $5\%$.  The uncertainty on the hadronic energy scale of the LAr
  calorimeter is $2\%$ and is found to have only a small effect on
  the cross sections in the present analysis.

\item The efficiency of the cut on the logarithmic energy-weighted
  cluster radius, $r_{log}$, is known to a precision of $0.5\%, 1.5\%$
  and $3\%$ for $0.6 < y \leq 0.7$, $0.7 < y \leq 0.8$ and $0.8 < y
  \leq 0.9$, respectively.

\item The charge asymmetry in the lepton candidates from photoproduction 
background events of $0.98$ is
  known to $4\%$ precision \cite{ref20}.

\item The RAPGAP MC is weighted by the ratio of $\sigma_r^D / F_2^D$
  in order to describe the data at high~$y$. The 
associated uncertainty 
is evaluated by replacing $F_L^D$ in 
the expression used for $\sigma_r^D$ 
in the reweighting procedure (equation~\ref{eqn:sigma})
by either $0.5 \cdot F_L^D$ or $1.5 \cdot F_L^D$.

\item The kinematic dependences of the model used to correct the data
  are generally well constrained from previous measurements. The
  uncertainties on the $t$, $\beta$ and $\xpom$ dependences are
  evaluated by weighting the generator-level kinematics by $e^{\pm
    t}$, $\beta^{\pm 0.05}$, $(1-\beta)^{\pm 0.05}$ and
  $(1/\xpom)^{0.05}$.  The effects of weighting in $t$ and $(1 -
  \beta)$ are found to have a negligible effect on the 
measured cross sections.

\item The uncertainty due to the resonant contributions modelled by
  DIFFVM is evaluated by calculating the change in acceptance when
  including this contribution in the simulation or not.

\item The non-diffractive DIS and QED-Compton backgrounds are modelled
  using MC simulations and are statistically subtracted from the
  data. The non-diffractive DIS background has a negligible effect in
  this analysis except at the highest
  $\xpom$. The QED-Compton events are only relevant for $M_X
  \rightarrow 0$. The normalisations of these backgrounds are
  controlled at the level of $100\%$ and $30\%$, respectively.

\item The corrections due to the finite measurement intervals 
(bin-centre corrections) are
  subject to an uncertainty, which is evaluated from the change in
  these corrections when this procedure is carried out using the H1
  2006 DPDF Fit~A and Fit~B parameterisations of the reduced cross
  section. The uncertainty is very small except at large $\beta$,
  where the shape of $\sigma_r^D$ is not well constrained, and at low
  $\beta$, corresponding to high~$y$.

\end{itemize}

Sources of experimental uncertainty which lead to systematic errors
which are not correlated between data at different $E_p$ values are the
statistical errors of the MC simulations and the following.

\begin{itemize}
\item The vertex reconstruction efficiency of the CJC is controlled to the
level of $2\%$ for $\xpom > 10^{-3}$ and $10\%$ for $10^{-4} < \xpom <
10^{-3}$.

\item The trigger efficiency is $\gsim 99 \%$ and measured with a precision of
  $1\%$ using independently-triggered data.

\item The uncertainty in the efficiency of linking a track to a SpaCal cluster is $1.5\%$.

\item The uncertainty on the efficiency of the forward detector selection
for rejecting proton dissociative events is $0.5\%$ \cite{ref43}.
\end{itemize}

The model dependent uncertainties on the factors applied in 
correcting the measurements to the $M_Y$ and $t$ ranges given in equation
$\ref{eqn:myt}$ are evaluated using the method described in \cite{ref6}.
The resulting normalisation uncertainities
are $7\%$ for all beam energies, dominated by the uncertainty on the
ratio of proton elastic to proton dissociative cross sections. This is
added in quadrature to the uncertainty of $3(4)\%$ on the luminosity
measurement to obtain the total normalisation uncertainty of
$7.6(8.1)\%$ for the $920\gev$ (LME) data.

A full decomposition of the systematic errors on the measured cross
sections is given in tables $\ref{tab:2}$, $\ref{tab:3}$ and
$\ref{tab:4}$.  Correlated sources of uncertainty that are 
always smaller than $2\%$
and are never the dominant correlated source in a single bin 
are omitted. For the LME data, the precision of the cross section
measurements is statistically limited in the 
region of greatest sensitivity to $F_L^D$ at high $y$.
Elsewhere in the LME data, the systematic errors are of
similar size to the statistical errors. The precision of $4\%$
reached in the best-measured regions for the $920\gev$ data is the
highest accuracy achieved in 
H1 measurements of $\sigma_r^D$ to date. The $920\gev$ data
are limited by the systematic
uncertainties throughout the measured range, the dominant source of 
systematic uncertainty varying
with the kinematics. The largest correlated uncertainty at low $\xpom$
comes from the modelling of the LAr noise, with the vector meson
simulation also playing an important role. At low $\beta$ (high~$y$),
where $F_L^D$ is measured, the largest sources of uncertainty are the
photoproduction background subtraction, the efficiency of the
$r_{log}$ cut and the model dependence arising from the $F_L^D$
treatment in the MC simulation. The uncertainty arising from imperfect
knowledge of the bin-centre corrections can also be large, typically
at large $\beta$, low $\xpom$ or low $Q^2$.

\subsection{Extraction of {\boldmath $F_L^D$}}
\label{subsec:FLD}

The separation of $F_2^D$ and $F_L^D$ follows a similar procedure to
that which was used to extract their inclusive counterparts $F_2$ and
$F_L$ \cite{ref20}. The diffractive reduced cross section is
integrated over the $M_Y$ and $t$ ranges given in equation
$\ref{eqn:myt}$. The uncertainty on correcting an individual dataset
to that range is large ($7\%$) but strongly correlated between
datasets. The residual difference in normalisation between the three
datasets after all corrections is determined from comparisons 
of $\sigma_r^D$ at low
$y$ to be $2\%$. In order to extract $F_L^D$ optimally, the cross
sections are normalised to the H1 2006 DPDF Fit~B result in a range
where the sensitivity to $F_L^D$ is minimal, but the statistical
precision and kinematic overlap of the data is still sufficient. Data
in the range $Q^2 > 7\gevsq$, $\xpom = 0.003$ and $y < 0.38$ ($0.3$
and $0.3$) for the $460$ ($575$ and $920\gev$) datasets are used,
yielding normalisation factors of $0.97$, $0.99$ and $0.97$,
respectively. As the published data at $820\gev$ were included in the
analysis of the data used as input to the H1 2006 DPDF Fit~B, they are
already consistently normalised.

Following this normalisation procedure, the diffractive longitudinal
structure function $F_L^D$ can be extracted directly from the slope of
$\sigma_r^D$ as a function of $y^2/Y_+$ for each set of $Q^2$, $\xpom$
and $\beta$ values. A linear fit is performed, taking only the
statistical errors, $\delta_{stat}$, into account in order to
calculate the statistical uncertainty on $F_L^D$. The fit is repeated,
adding the statistical and uncorrelated errors in quadrature,
$\delta_{stat+unc}$, to calculate the measured value of $F_L^D$ and
the sum of its combined statistical and uncorrelated errors. For each
correlated systematic error source, each of the cross section points
is adjusted according to the positive and negative shifts\footnote{ In
  fits which include the published $820\gev$ data, a more conservative approach
  is used whereby the $820\gev$ data remain fixed. This results in a
  larger variation in the slope of $\sigma_r^D$ as a function of
  $y^2/Y_+$ with a correspondingly larger uncertainty on $F_L^D$.}
and the fit is repeated using $\delta_{stat+unc}$ for the errors on
the cross section points. The error on $F_L^D$ is taken as half of the
difference between fits to the positive and negative shifted data
points. All of these correlated errors are added in quadrature with
$\delta_{stat+unc}$ to give the total error on $F_L^D$. 
The normalisation uncertainty on the value of $F_L^D$ is set by the
normalisation uncertainty on the cross section measurements and is
therefore $8.1\%$.

As only
bin-averaged cross sections are available at the highest $\beta >
0.9$, $F_L^D$ is not extracted in that region.

\subsection{Extraction of {\boldmath $R^D$} and the ratio {\boldmath $R^D / R$}}

The photoabsorption ratio for diffraction, $R^D = F_L^D /(F_2^D -
F_L^D)$, is extracted from linear fits to the data by reparametrising
equation $\ref{eqn:sigma}$ such that $R^D$ and $(F_2^D - F_L^D)$ 
become the free 
parameters of the fit:
\begin{equation}
\sigma_r^D = (F_2^D - F_L^D) + R^D \cdot (F_2^D - F_L^D) \cdot (1 - y^2/Y_+).
\end{equation}
The error on $R^D$ is calculated in the same way as for $F_L^D$,
detailed in section $\ref{subsec:FLD}$. The normalisation uncertainty
cancels in this ratio.

In order to calculate the ratio of $R^D$ to its inclusive counterpart
$R = F_L/(F_2 - F_L)$, the value of $R$ is extracted from 
the present data
using a similar procedure to that used for $R^D$ described above. Only
data with $Q^2 > 7\gevsq$ are used, where inclusive measurements are
made at all beam energies in this analysis. The statistical
correlations between the inclusive and diffractive measurements are
neglected and the systematic errors are assumed to be dominated by the
error on $R^D$. Similarly to $R^D$, there is no normalisation 
uncertainty on the ratio $R^D / R$.

\section{Results}

\begin{figure}[htbp]
  \begin{center}
    \includegraphics[width=1.0\textwidth]{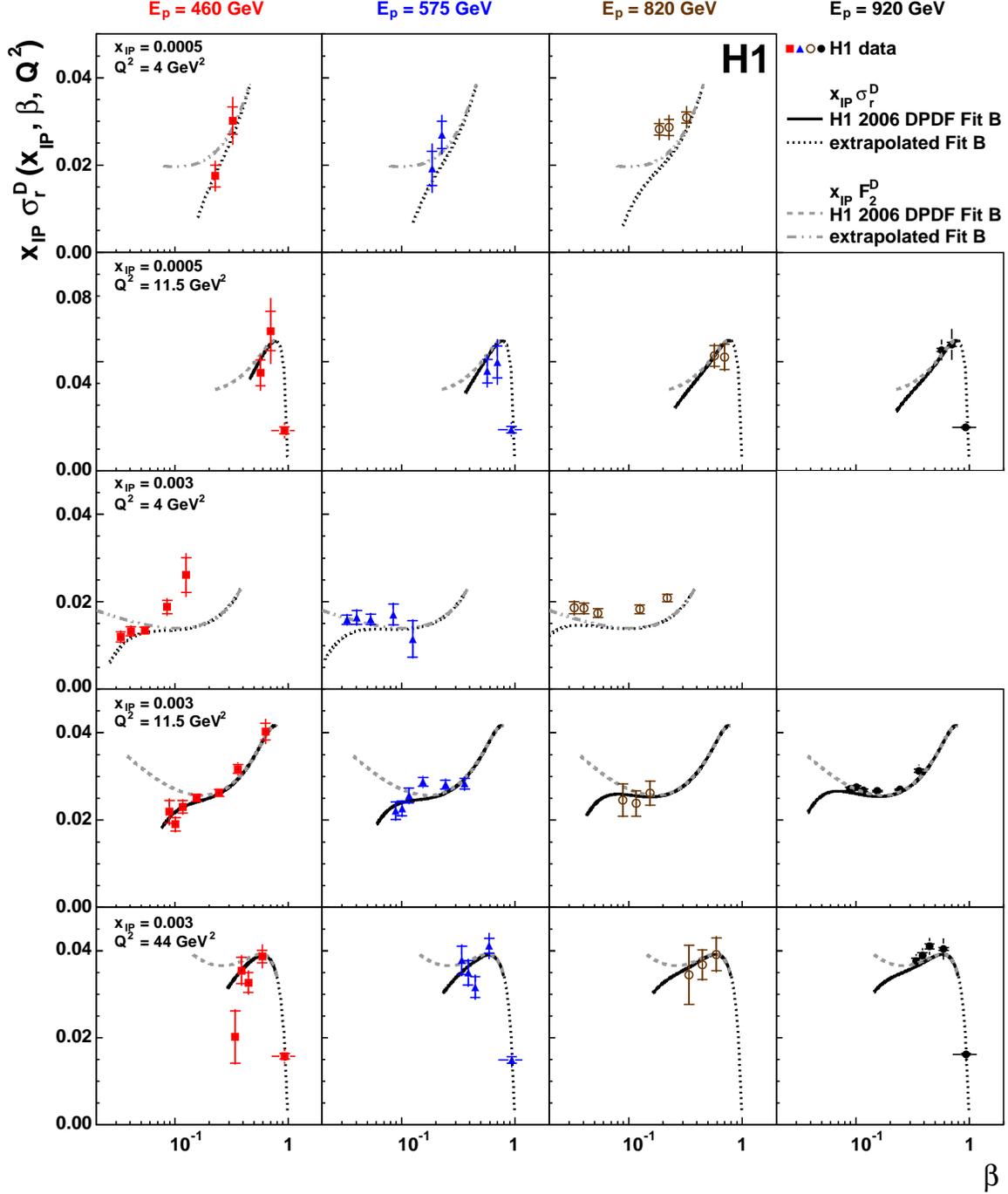}
  \end{center}
  \caption{The diffractive reduced cross section $\sigma_r^D$
    multiplied by $\xpom$ as a function of $\beta$ at fixed $Q^2$ and
    $\xpom$ for (from left to right) the $460\gev$, $575\gev$,
    $820\gev$ and $920\gev$ datasets.  The data are
    compared with the predictions of H1 2006 DPDF Fit~B (solid line),
    which is indicated as dotted beyond the range of validity of the
    fit. The dashed and dashed-dotted lines represent the contribution 
    of $F_2^D$, which
    is the same for each beam energy. The inner error bars represent the
    statistical errors on the measurement, the outer error bars
    represent the statistical and total systematic uncertainties added in
    quadrature. The normalisation uncertainties of $7.6(8.1)\%$ for
    the $920(460,575)\gev$ data are not shown.}
\label{fig:5}
\end{figure}

The measured diffractive reduced cross section values and their errors
are given in tables $\ref{tab:2}$, $\ref{tab:3}$ and
$\ref{tab:4}$. Figure $\ref{fig:5}$ shows the reduced cross section as
a function of $\beta$ at fixed $\xpom$ and $Q^2$ for the LME,
$820\gev$ and $920\gev$ datasets. Also shown is the prediction of H1
2006 DPDF Fit~B, which in general describes the data well at $Q^2 \geq
11.5\gevsq$. Deviations of the measured cross sections from the $F_2^D$
predictions at low $\beta$ are evident in the LME data, where the
highest $y$ values are accessed, notably at $Q^2 = 11.5\gevsq$ and
$\xpom = 0.003$. This shows the sensitivity of the LME data to
$F_L^D$. The extrapolation to lower $Q^2$ of H1 2006 DPDF Fit~B, which
only included data with $Q^2 \geq 8.5\gevsq$, is also compared with the
$Q^2 = 4\gevsq$ data. The fit is known to significantly undershoot
the published $820\gev$ data in this region \cite{ref6}, an observation which
is reproduced for the new measurements.

The new data at $\xpom = 0.0005$, $Q^2 = 11.5\gevsq$ and
$\xpom=0.003$, $Q^2=44\gevsq$ include the highest $\beta$ measurements
obtained by H1 to date. They are in remarkably good agreement with the
extrapolation of H1 2006 DPDF Fit~B and support the hypothesis that
$\sigma_r^D \rightarrow 0$ as $\beta \rightarrow 1$. There is thus no
evidence in this region for a large higher twist $F_L^D$ contribution
\cite{ref4,ref26,ref27}.

\begin{figure}[htbp]
  \begin{center}
    \includegraphics[width=1.0\textwidth]{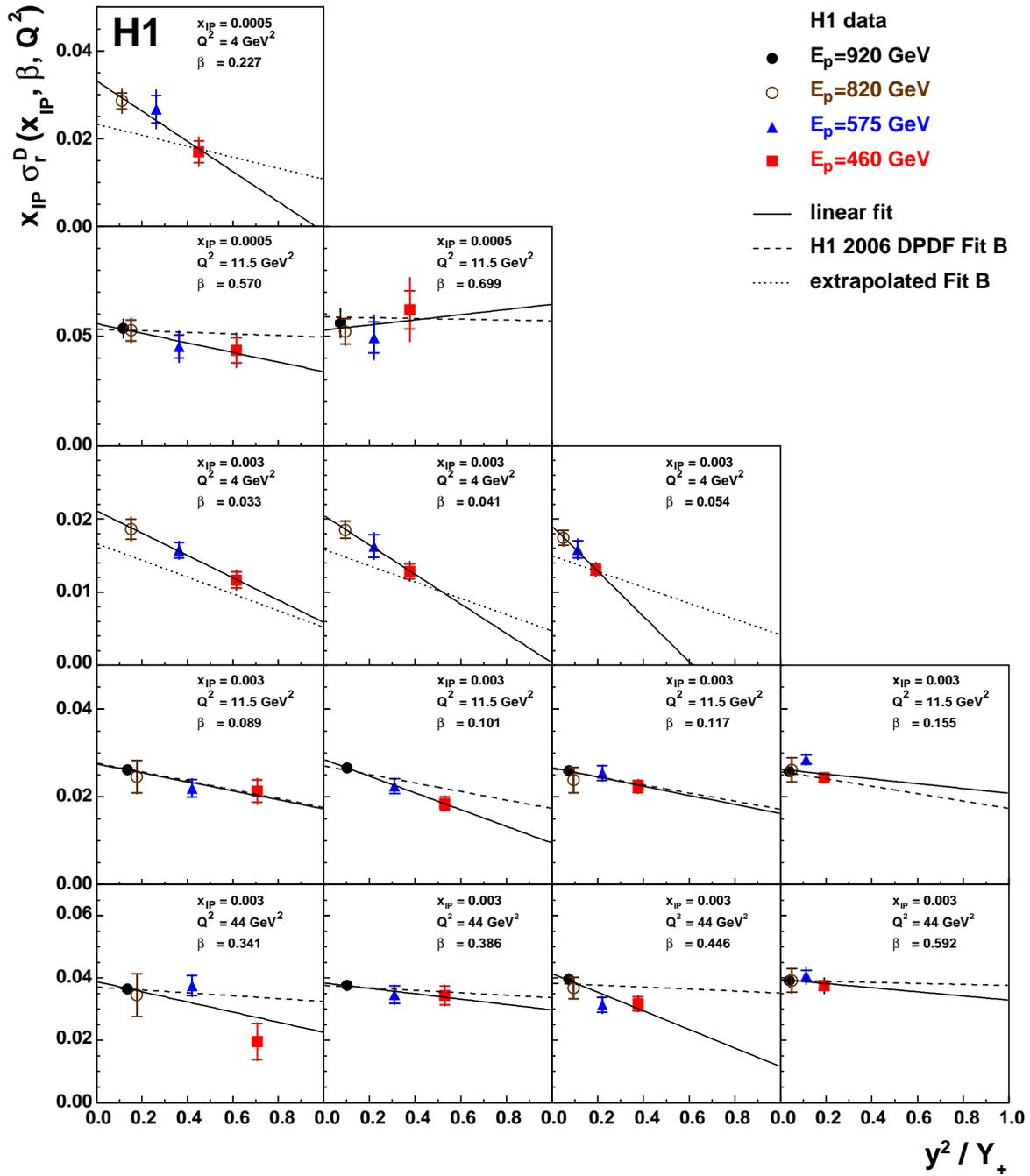}
  \end{center}
\vspace*{-.3cm}
  \caption{The diffractive reduced cross section $\sigma_r^D$
    multiplied by $\xpom$ as a function of $y^2/Y_+$ at fixed $Q^2$,
    $\xpom$ and $\beta$.  The inner error bars represent the
    statistical uncertainties on the measurement, the outer error bars
    represent the statistical and total systematic 
uncertainties added in
    quadrature. The normalisation uncertainty is not shown. Up to four beam
    energies are shown, where the lowest $y^2/Y_+$ point is given by
    the $820\gev$ data for $Q^2 = 4\gevsq$ and by the $920\gev$ data
    at higher $Q^2$. The linear fits to the data are also shown as a
    solid line, the slope of which gives the value of $F_L^D$. The
    predictions and extrapolated predictions of H1 2006 DPDF Fit~B are
    shown as dashed and dotted lines, respectively.}
\label{fig:6}
\end{figure}

The extraction of $F_L^D$ via linear fits to the $y^2/Y_+$ dependence
of the reduced cross section at different beam energies and fixed $Q^2$,
$\beta$
and $\xpom$ is shown in figure $\ref{fig:6}$. The largest lever arm in
$y^2/Y_+$, and therefore the highest sensitivity to $F_L^D$, is at the
lowest $\beta$. The data are consistent with a linear dependence of
$\sigma_r^D$ on $y^2/Y_+$, with a significant tendency for
$\sigma_r^D$ to decrease as $y^2/Y_+$ increases for most $Q^2$,
$\xpom$ and $\beta$ values. The values of $F_L^D$ and their errors are
given in table $\ref{tab:5}$.

\begin{figure}[htbp]
  \begin{center}

    \includegraphics[width=0.8\textwidth]{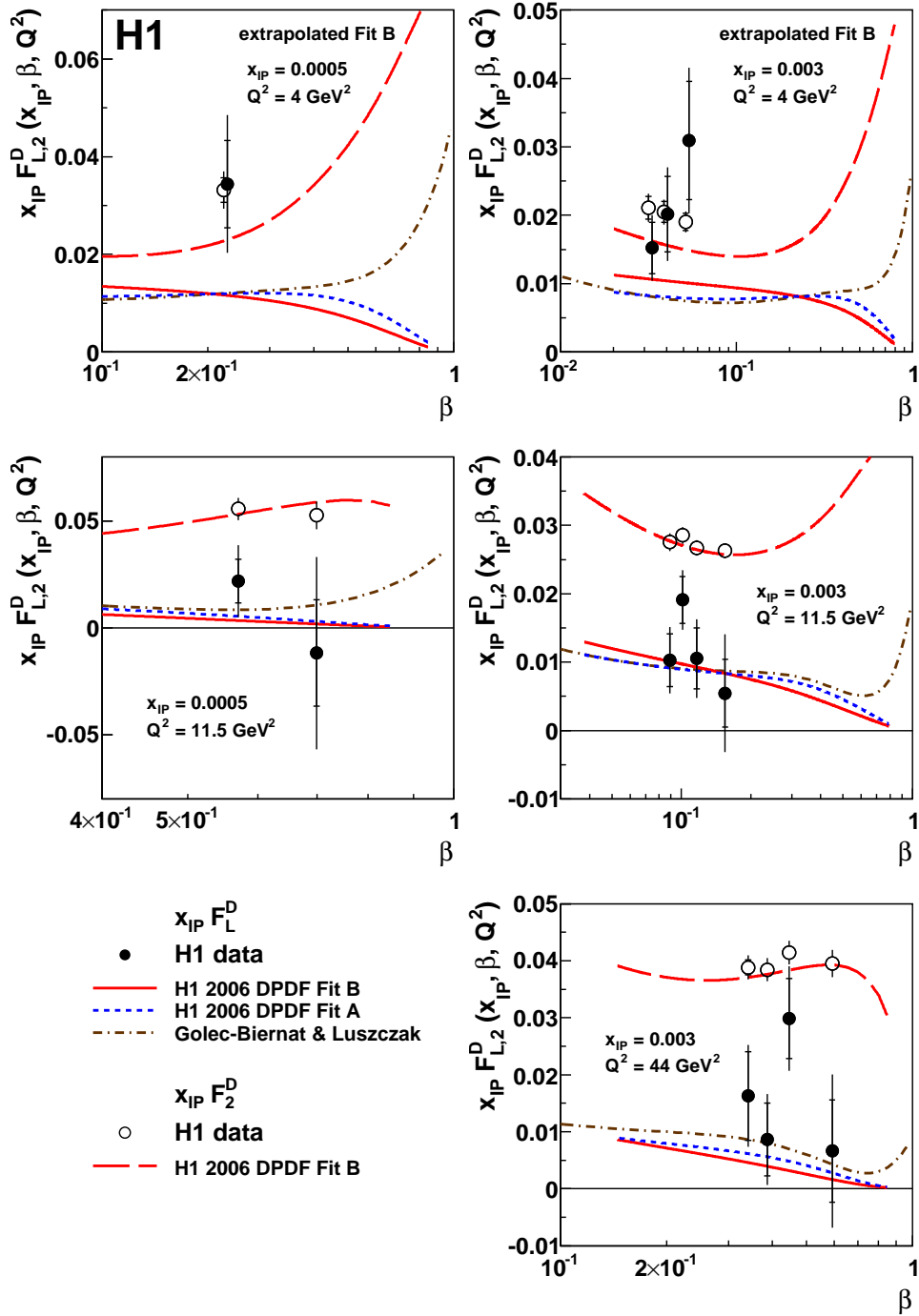}
  \end{center}
  \caption{The diffractive structure functions $F_L^D$ and $F_2^D$
    multiplied by $\xpom$ as a function of $\beta$ at fixed $Q^2$ and
    $\xpom$.  The $F_L^D$ data are shown as filled points, compared
    with the predictions of H1 2006 DPDF Fit~A (dashed line), Fit~B
    (solid line) and the Golec-Biernat and \L uszczak model (dashed
    and dotted line).  The measurements of $F_2^D$ (open points) are
    compared with the prediction of H1 2006 DPDF Fit~B (long dashed
    line). The inner error bars represent the statistical 
uncertainties on the
    measurement, the outer error bars represent the statistical and
    total systematic uncertainties added in quadrature. The normalisation
    uncertainty of $8.1\%$ is not shown.}
\label{fig:7}
\end{figure}

The measurements of $F_L^D$, at fixed values of $Q^2$ and $\xpom$, are
shown as a function of $\beta$
in figure $\ref{fig:7}$. Significantly non-zero measurements of
$F_L^D$ are made for all values of $Q^2$ and $\xpom$ and five $F_L^D$
points are greater than zero by more than $3\sigma$. The data are
compared with the predictions of the H1 DPDF Fits A and B \cite{ref6}
and with the Golec-Biernat \& \L uszczak model \cite{ref29} (section
$\ref{models}$). 
All three models are consistent with
the data, although there is a tendency for the measurements to lie
above the predictions. 
Although the prediction of \cite{ref29} lies
significantly above both Fit~A and Fit~B at large $\beta$, 
the experimental precision is insufficient to
distinguish between the models. 
The
measured values of $F_2^D$ are
also shown in figure $\ref{fig:7}$.  
The $F_2^D$ measurements agree well with
the predictions of H1 DPDF Fit~B for $Q^2 \geq 11.5\gevsq$. Within the
uncertainties, all measurements are consistent with the hypothesis
that $0 < F_L^D < F_2^D$.

\begin{figure}[htbp]
  \begin{center}
    \includegraphics[width=1.0\textwidth]{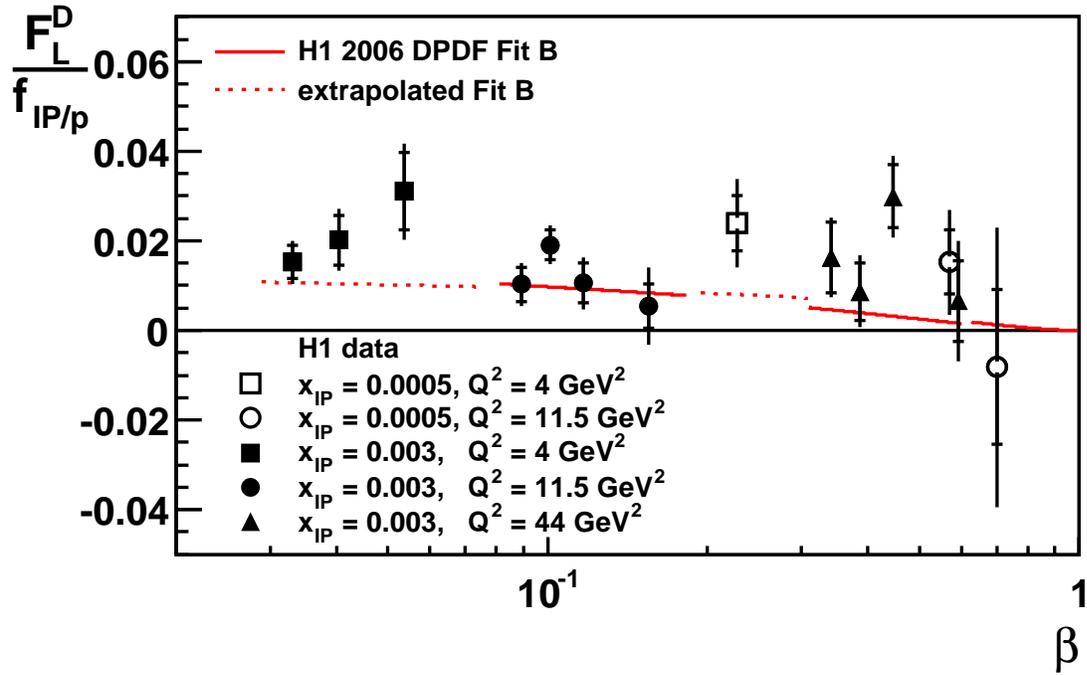}
  \end{center}
  \caption{The diffractive longitudinal structure function $F_L^D$,
    divided by a parametrisation of the $\xpom$ dependence
    of the reduced cross section
    $f_{I\!\!P/p}$ \cite{ref6}, as a function of $\beta$ at the 
    indicated values of
    $Q^2$ and $\xpom$.  The data are compared with the
    predictions of H1 2006 DPDF Fit~B (red line), which is indicated
    as dashed beyond the range of validity of the fit. The inner error
    bars represent the statistical uncertainties on the measurement, the outer
    error bars represent the statistical and total systematic 
uncertainties
    added in quadrature. The normalisation uncertainty of $8.1\%$ is
    not shown.}
\label{fig:8}
\end{figure}

A summary of the $F_L^D$ measurements is given in figure
$\ref{fig:8}$, where the data points from all five $Q^2$ and $\xpom$ values
are shown as a function of $\beta$ and compared with the H1 2006 DPDF
Fit~B prediction. In order to remove the significant dependence on
$\xpom$, the $F_L^D$ points have been divided by a factor
$f_{I\!P/p}$, taken from \cite{ref6}, which expresses the measured $\xpom$
dependence of the data, assuming proton vertex factorisation. The
remaining discontinuities in the prediction are due to its $Q^2$
dependence. The $F_L^D$ data cover a large range in longitudinal
fractional momentum $0.033 < \beta < 0.7$ and are compatible with the
predicted slow decrease with increasing $\beta$. The data have a
tendency to lie above the prediction although the precision is
limited. The most significantly positive $F_L^D$ measurements lie in
the region $\beta < 0.5$, which contrasts with models of
diffraction such as \cite{ref4,ref26}, which do not include leading twist
contributions from longitudinally polarised photons.

\begin{figure}[htbp]
  \begin{center}
    \includegraphics[width=1.0\textwidth]{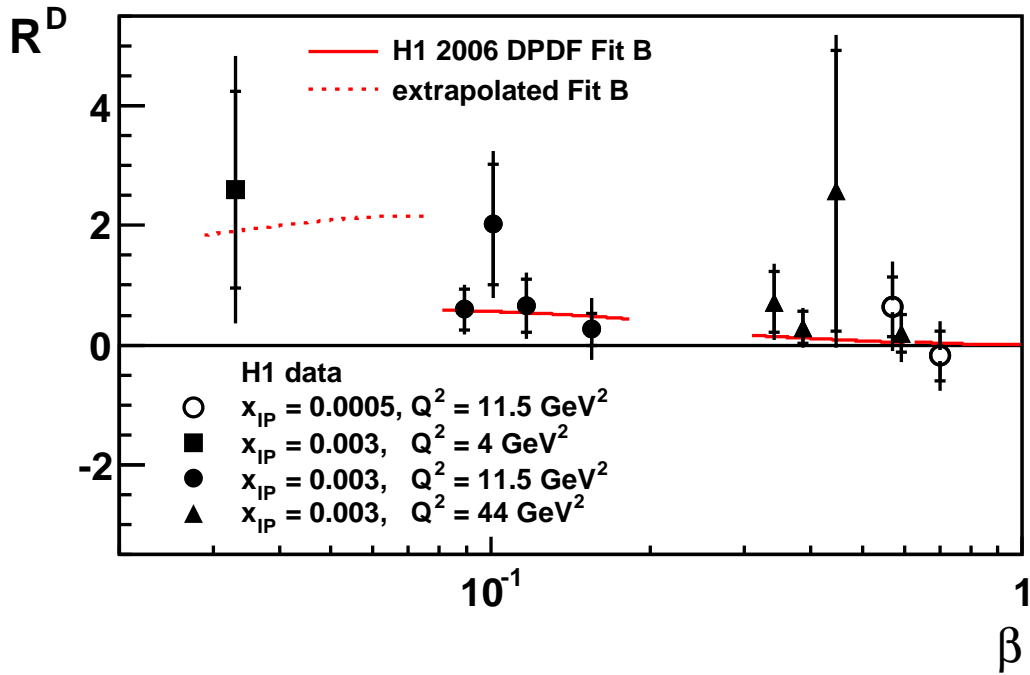}
  \end{center}
  \caption{ The ratio $R^D$  of cross sections for longitudinally 
to transversely polarised photons, 
    as a function of $\beta$ at the indicated values of $\xpom$ and $Q^2$. The
    data are compared with the predictions of H1 2006 DPDF Fit~B,
    indicated as dashed beyond the range of validity of the fit. The
    inner error bars represent the statistical uncertainties on the
    measurement, the outer error bars represent the statistical and
    total systematic uncertainties added in quadrature.}
\label{fig:9}
\end{figure}

The measurement of $R^D$ is shown as a function of $\beta$ in figure
$\ref{fig:9}$. Data with $|R^D| > 50$ and a relative uncertainty
larger than $100\%$ are not shown. The data are compatible with the
prediction based on H1 2006 DPDF Fit~B, though they are also
consistent with other models. The data at $Q^2 = 11.5\gevsq$ indicate
that the longitudinally and transversely polarised photon cross
sections are of the same order of magnitude ($R^D \sim 1$ and $F_2^D
\sim 2F_L^D$). At $Q^2 = 44\gevsq$, where larger $\beta$ values are
accessed, there is a tendency for the data to lie above the
prediction, which tends to zero as $\beta \rightarrow 1$. There is no
evidence for the steep rise in $R^D$ which might be expected at large
$\beta$ if configurations similar to vector meson electroproduction
were dominant in this region.  The values of $R^D$ and their errors
are given in table $\ref{tab:6}$.

\begin{figure}[htbp]
  \begin{center}
    \includegraphics[width=0.6\textwidth]{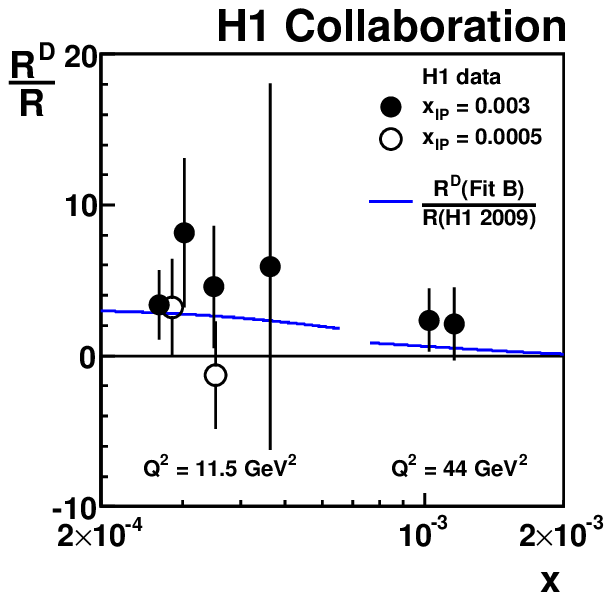}
  \end{center}
  \caption{ The ratio of $R^D/R$ as a function of $x$ at 
the indicated values of $Q^2$
    and $\xpom$. The data are compared with the predicted ratio using
    H1 2006 DPDF Fit~B / H1 PDF 2009 (solid line). The error bars
    represent the statistical and systematic uncertainties added in
    quadrature.}
\label{fig:10}
\end{figure}

The relative importance in inclusive and diffractive scattering of the
longitudinally polarised photon cross section compared with its
transverse counterpart is investigated via the ratio $R^D/R$, shown
as a function of $x$ in figure $\ref{fig:10}$. Only data with $Q^2 >
7\gevsq$, where a measurement of $R$ is possible in this analysis, are
used. Data with $|R^D/R| > 20$ and a relative uncertainty greater than
$100\%$ are not shown. The ratio data suggest that the longitudinally
polarised photon contribution plays a larger role in the diffractive
than the inclusive case. Averaged over all data, $R^D/R = 2.8 \pm
1.1$. The data are well reproduced by the ratio of predictions from H1
2006 DPDF Fit~B and an H1 fit to inclusive DIS data, H1 PDF 2009
\cite{ref44}. At high $Q^2$, corresponding to high $x$ and therefore
$\beta$, the prediction decreases towards zero as $x \rightarrow
1$. The data are consistent with such a decrease with increasing
$\beta$ within large experimental uncertainties.

\section{Conclusions}

First measurements of the diffractive reduced cross section at
centre-of-mass energies $\sqrt{s}$ of $225$ and $252\gev$ are
presented, together with a precise measurement at $\sqrt{s}$ of
$319\gev$. The reduced cross section is measured in the range of
photon virtualities $4.0 \leq Q^2 \leq 44.0\gevsq$ and 
of the longitudinal
momentum fraction of the diffractive exchange $5 \cdot 10^{-4} \leq \xpom \leq 3 \cdot
10^{-3}$. The reduced cross section measurements agree well with
predictions derived from leading twist NLO QCD fits to previous H1
data 
throughout the kinematic range. The data at high and
medium inelasticity $y$ are used to extract the first
measurement of the longitudinal diffractive structure function
$F_L^D$. 
There is a tendency for the predictions to lie below the $F_L^D$ data, but 
the data are compatible with H1 2006 DPDF Fit~A and
Fit~B as well as with a model which includes a higher twist
contribution at high $\beta$, based on a colour dipole approach.
 The procedure also allows a simultaneous extraction of
$F_2^D$, independently of assumptions made on $F_L^D$, in the same
kinematic range. The $F_2^D$ measurements agree well with the
predictions of H1 DPDF Fit~B for $Q^2 \geq 11.5\gevsq$. Within the
uncertainties, all measurements are consistent with the expectation
that $0 < F_L^D < F_2^D$.

The ratio $R^D$ of diffractive cross sections for longitudinally to
transversely polarised photons is measured in the same kinematic range
as $F_L^D$. At fixed $Q^2$ and $\xpom$, this ratio is relatively flat
as a function of $\beta$ and suggests that the cross sections for the
two polarisation states of the photon are of comparable size. The
ratio of $R^D$ to its inclusive scattering counterpart, $R$, is
extracted in the region $Q^2 \geq 11.5\gevsq$. The $R^D/R$ data
indicate that the longitudinally polarised photon cross section plays
a larger role in the diffractive than in the inclusive case. The $R^D$
and $R^D/R$ measurements are well reproduced by the predictions based
on H1 2006 DPDF Fit~B and the H1 PDF 2009 inclusive PDF set.

\section*{Acknowledgements}

We are grateful to the HERA machine group whose outstanding efforts
have made this experiment possible.  We thank the engineers and
technicians for their work in constructing and maintaining the H1
detector, our funding agencies for financial support, the DESY
technical staff for continual assistance and the DESY directorate for
support and for the hospitality which they extend to the non DESY
members of the collaboration. We also thank K. Golec-Biernat for 
providing us with the Golec-Biernat \& \L uszczak model predictions.

\pagebreak

\begin{landscape} 
\begin{table} 
\begin{footnotesize} 
\begin{center}
\begin{tabular}{|c | c | c || c | c | c | c | c || c | c | c | c | c | c | c | c | c | c | c | c |} 
  \hline 
  $\xpom$ & $Q^2$ & $\beta$ & $\xpom\sigma_r^D$ & $\delta_{stat}$ & $\delta_{unc}$ & $\delta_{cor}$ & $\delta_{tot}$ & $\delta_{ele}$ & $\delta_{\theta}$ & $\delta_{noi}$ & $\delta_{spa}$ & $\delta_{r_{log}}$ & $\delta_{asy}$ & $\delta_{mod}$ & $\delta_{\beta}$ & $\delta_{\xpom}$ & $\delta_{vm}$ & $\delta_{com}$ & $\delta_{bcc}$ \\
  & $[\gevsq]$ & & & $[\%]$ & $[\%]$ & $[\%]$ & $[\%]$ & $[\%]$ & $[\%]$ & $[\%]$ & $[\%]$ & $[\%]$ & $[\%]$ & $[\%]$ & $[\%]$ & $[\%]$ & $[\%]$ & $[\%]$ & $[\%]$\\ \hline\hline
$0.0005 $&$  4.0 $&$ 0.227 $&$ 0.0175 $&$  14.2 $&$  17.4 $&$  15.2 $&$  20.2 $&$  2.4 $&$ -0.7 $&$  6.3 $&$  3.9 $&$  3.7 $&$ -0.9 $&$  0.3 $&$  2.5 $&$  0.8 $&$  8.2 $&$ -0.4 $&$  7.1 $\\
$0.0005 $&$  4.0 $&$ 0.323 $&$ 0.0302 $&$  10.4 $&$  11.6 $&$   9.5 $&$  18.3 $&$  2.0 $&$  1.4 $&$  6.6 $&$  3.1 $&$ -0.0 $&$  0.0 $&$  0.7 $&$  0.9 $&$  0.0 $&$  4.7 $&$  0.0 $&$ -2.0 $\\
$0.0005 $&$ 11.5 $&$ 0.570 $&$ 0.0448 $&$  13.2 $&$  11.0 $&$   8.7 $&$  18.4 $&$  1.8 $&$  0.2 $&$  7.0 $&$  1.3 $&$  0.3 $&$ -2.4 $&$ -2.6 $&$  0.4 $&$  1.3 $&$  0.8 $&$ -8.2 $&$  1.0 $\\
$0.0005 $&$ 11.5 $&$ 0.699 $&$ 0.0640 $&$  14.0 $&$  12.1 $&$  14.8 $&$  23.7 $&$  1.1 $&$  1.3 $&$ 12.5 $&$  1.6 $&$  0.9 $&$ -0.5 $&$  0.9 $&$  1.0 $&$  0.0 $&$  6.7 $&$ -0.5 $&$  3.8 $\\
$0.0005 $&$ 11.5 $&$ 0.755-1.0 $&$ 0.0185 $&$   8.8 $&$  10.7 $&$  14.9 $&$  20.4 $&$  2.2 $&$  0.6 $&$ 12.6 $&$  1.2 $&$  0.3 $&$  0.0 $&$  1.8 $&$  0.7 $&$  0.0 $&$  4.5 $&$  0.0 $ & $0.0$\\
$ 0.003 $&$  4.0 $&$ 0.033 $&$ 0.0120 $&$   9.6 $&$   5.1 $&$   6.3 $&$  12.2 $&$  1.7 $&$ -0.6 $&$ -0.0 $&$  0.1 $&$ -5.7 $&$ -2.4 $&$ -2.9 $&$ -0.2 $&$  3.1 $&$ -0.6 $&$ -0.1 $&$ -0.8 $\\
$ 0.003 $&$  4.0 $&$ 0.041 $&$ 0.0132 $&$   8.0 $&$   4.9 $&$   5.3 $&$  10.8 $&$  0.9 $&$ -1.0 $&$ -0.6 $&$ -0.0 $&$ -4.1 $&$ -0.5 $&$ -1.4 $&$ -1.2 $&$  0.4 $&$ -0.3 $&$  0.0 $&$ -0.6 $\\
$ 0.003 $&$  4.0 $&$ 0.054 $&$ 0.0135 $&$   5.6 $&$   3.7 $&$   4.0 $&$   7.8 $&$  1.9 $&$ -0.7 $&$ -0.6 $&$ -0.3 $&$ -1.5 $&$  0.0 $&$ -0.2 $&$ -1.8 $&$  0.0 $&$ -0.4 $&$ -0.1 $&$  0.6 $\\
$ 0.003 $&$  4.0 $&$ 0.085 $&$ 0.0188 $&$   8.3 $&$   4.9 $&$   5.4 $&$  11.0 $&$  0.5 $&$ -1.8 $&$ -1.9 $&$  0.0 $&$ -0.6 $&$  0.0 $&$  0.1 $&$ -2.0 $&$  0.0 $&$  0.7 $&$  0.0 $&$  3.0 $\\
$ 0.003 $&$  4.0 $&$ 0.125 $&$ 0.0261 $&$  15.0 $&$   8.8 $&$   7.9 $&$  19.1 $&$ -1.4 $&$ -3.5 $&$ -1.4 $&$  1.5 $&$ -0.2 $&$  0.0 $&$  0.0 $&$ -0.9 $&$  0.0 $&$  1.1 $&$  0.0 $&$  6.3 $\\
$ 0.003 $&$ 11.5 $&$ 0.089 $&$ 0.0219 $&$  11.9 $&$   4.2 $&$   6.6 $&$  14.3 $&$  2.1 $&$  0.8 $&$ -1.7 $&$ -0.5 $&$ -1.9 $&$ -3.2 $&$ -3.0 $&$ -0.8 $&$  2.9 $&$ -0.8 $&$ -0.5 $&$  1.2 $\\
$ 0.003 $&$ 11.5 $&$ 0.101 $&$ 0.0190 $&$   8.3 $&$   3.5 $&$   3.9 $&$   9.8 $&$  0.9 $&$  0.6 $&$ -0.5 $&$ -0.1 $&$ -1.1 $&$ -1.6 $&$ -1.3 $&$ -0.6 $&$  1.1 $&$ -0.9 $&$  0.0 $&$  0.9 $\\
$ 0.003 $&$ 11.5 $&$ 0.117 $&$ 0.0230 $&$   6.3 $&$   3.3 $&$   3.1 $&$   7.8 $&$  0.7 $&$  1.2 $&$ -0.3 $&$  0.2 $&$ -0.3 $&$ -0.5 $&$ -0.7 $&$ -0.2 $&$  0.3 $&$ -0.6 $&$  0.0 $&$  0.8 $\\
$ 0.003 $&$ 11.5 $&$ 0.155 $&$ 0.0251 $&$   3.2 $&$   2.5 $&$   2.6 $&$   4.8 $&$  1.0 $&$  0.9 $&$  0.2 $&$  0.2 $&$ -0.0 $&$  0.0 $&$ -0.3 $&$ -0.0 $&$  0.0 $&$  0.1 $&$ -0.0 $&$  0.7 $\\
$ 0.003 $&$ 11.5 $&$ 0.244 $&$ 0.0262 $&$   3.0 $&$   2.4 $&$   2.8 $&$   4.7 $&$  0.7 $&$  0.9 $&$  0.5 $&$ -0.4 $&$ -0.1 $&$  0.0 $&$ -0.1 $&$ -0.3 $&$  0.0 $&$  0.8 $&$ -0.2 $&$  0.5 $\\
$ 0.003 $&$ 11.5 $&$ 0.361 $&$ 0.0317 $&$   3.1 $&$   2.5 $&$   2.6 $&$   4.8 $&$  0.3 $&$  1.1 $&$  1.0 $&$  0.7 $&$  0.0 $&$  0.0 $&$  0.0 $&$ -0.1 $&$  0.0 $&$ -0.2 $&$ -0.1 $&$ -0.0 $\\
$ 0.003 $&$ 11.5 $&$ 0.631 $&$ 0.0403 $&$   4.7 $&$   3.0 $&$   4.6 $&$   7.2 $&$ -3.4 $&$  1.7 $&$  0.4 $&$  0.5 $&$  0.0 $&$  0.0 $&$ -0.1 $&$ -0.2 $&$  0.0 $&$  0.0 $&$ -0.1 $&$  1.1 $\\
$ 0.003 $&$ 44.0 $&$ 0.341 $&$ 0.0202 $&$  29.8 $&$   8.0 $&$   7.0 $&$  31.7 $&$  2.7 $&$ -1.7 $&$ -1.7 $&$ -0.6 $&$  1.5 $&$ -3.2 $&$ -2.1 $&$  1.3 $&$  1.8 $&$  2.9 $&$  0.0 $&$ -0.7 $\\
$ 0.003 $&$ 44.0 $&$ 0.386 $&$ 0.0355 $&$   8.6 $&$   4.2 $&$   3.1 $&$  10.0 $&$  0.6 $&$  0.5 $&$ -0.3 $&$ -0.5 $&$  0.3 $&$ -1.6 $&$ -0.8 $&$  0.5 $&$  0.3 $&$  0.1 $&$ -0.4 $&$  0.7 $\\
$ 0.003 $&$ 44.0 $&$ 0.446 $&$ 0.0327 $&$   7.0 $&$   3.5 $&$   3.6 $&$   8.6 $&$  1.0 $&$  0.6 $&$  0.4 $&$ -0.1 $&$  0.2 $&$ -0.5 $&$ -0.3 $&$  0.3 $&$  0.2 $&$ -0.3 $&$ -0.1 $&$  2.3 $\\
$ 0.003 $&$ 44.0 $&$ 0.592 $&$ 0.0387 $&$   3.8 $&$   2.6 $&$   5.5 $&$   7.2 $&$  0.4 $&$  1.2 $&$  1.4 $&$  0.3 $&$  0.0 $&$  0.0 $&$  0.0 $&$  0.2 $&$  0.0 $&$  0.7 $&$ -1.0 $&$  4.8 $\\
$ 0.003 $&$ 44.0 $&$ 0.76-1.0 $&$ 0.0157 $&$   4.1 $&$   2.7 $&$   9.9 $&$  11.0 $&$ -0.2 $&$  1.6 $&$  2.1 $&$ -0.4 $&$ -0.0 $&$  0.0 $&$  0.0 $&$ -0.1 $&$  0.0 $&$ -0.7 $&$ -1.7 $& $0.0$\\
  \hline
\end{tabular}
\end{center} 
\end{footnotesize} 
\caption{
  The diffractive reduced cross section $\sigma_r^D$ at $\sqrt{s} = 225\gev$, 
  multiplied by $\xpom$, measured with the $460\gev$ data, at fixed values of 
  $\xpom$, $Q^2$ and $\beta$.  At the largest $\beta$, the bin-averaged cross section
  is given together with the lower and upper bin boundaries.  The statistical ($\delta_{stat}$), 
  uncorrelated ($\delta_{unc}$) and sum of all correlated ($\delta_{cor}$)
uncertainties are given 
  together with the total uncertainty ($\delta_{tot}$).  The other columns show the individual 
  correlated uncertainties, which are due to the positron energy scale ($\delta_{ele}$), the positron
  polar angle measurement ($\delta_{\theta}$), the LAr noise subtraction ($\delta_{noi}$), the 
  hadronic SpaCal energy scale ($\delta_{spa}$), the efficiency of the logarithmic energy-weighted
  cluster radius cut ($\delta_{r_{log}}$), the charge asymmetry of the photoproduction background 
  ($\delta_{asy}$), the model uncertainty due to the influence of $F_L^D$ ($\delta_{mod}$), the model
  uncertainty on the underlying $\beta$ and $\xpom$ distributions ($\delta_{\beta}$, $\delta_{\xpom}$), 
  the influence of resonant ($\delta_{vm}$) and QED Compton ($\delta_{Com}$) contributions and finally
  the parametrisation choice for the bin centre corrections ($\delta_{bcc}$). A minus sign indicates
  that a source is anti-correlated with a change in the cross section. All uncertainties are 
are given in per cent.
The normalisation uncertainty of $8.1\%$ is not included.
}
\label{tab:2}
\end{table} 

\begin{table} 
\begin{footnotesize} 
\begin{center}
\begin{tabular}{|c | c | c || c | c | c | c | c || c | c | c | c | c | c | c | c | c | c | c | c |} 
  \hline 
  $\xpom$ & $Q^2$ & $\beta$ & $\xpom\sigma_r^D$ & $\delta_{stat}$ & $\delta_{unc}$ & $\delta_{cor}$ & $\delta_{tot}$ & $\delta_{ele}$ & $\delta_{\theta}$ & $\delta_{noi}$ & $\delta_{spa}$ & $\delta_{r_{log}}$ & $\delta_{asy}$ & $\delta_{mod}$ & $\delta_{\beta}$ & $\delta_{\xpom}$ & $\delta_{vm}$ & $\delta_{com}$ & $\delta_{bcc}$ \\
  & $[\gevsq]$ & & & $[\%]$ & $[\%]$ & $[\%]$ & $[\%]$ & $[\%]$ & $[\%]$ & $[\%]$ & $[\%]$ & $[\%]$ & $[\%]$ & $[\%]$ & $[\%]$ & $[\%]$ & $[\%]$ & $[\%]$ & $[\%]$\\ \hline\hline
$0.0005 $&$  4.0 $&$ 0.186 $&$ 0.0192 $&$  20.2 $&$  17.1 $&$  16.4 $&$  29.4 $&$  1.5 $&$  0.5 $&$  6.7 $&$  4.5 $&$  3.3 $&$ -0.7 $&$  1.2 $&$  2.9 $&$  0.6 $&$  6.1 $&$  0.0 $&$ 14.0 $\\
$0.0005 $&$  4.0 $&$ 0.227 $&$ 0.0269 $&$  11.6 $&$  13.3 $&$  11.3 $&$  16.7 $&$  2.1 $&$ -1.0 $&$  5.0 $&$  3.8 $&$  0.6 $&$ -0.1 $&$  0.2 $&$  1.7 $&$  0.1 $&$  5.8 $&$  0.0 $&$  6.8 $\\
$0.0005 $&$ 11.5 $&$ 0.570 $&$ 0.0456 $&$  11.6 $&$  12.4 $&$  13.7 $&$  16.2 $&$  1.4 $&$  1.8 $&$  8.3 $&$  1.8 $&$  0.7 $&$ -0.5 $&$  0.2 $&$  1.0 $&$  0.2 $&$  6.8 $&$ -0.2 $&$  1.1 $\\
$0.0005 $&$ 11.5 $&$ 0.699 $&$ 0.0498 $&$  14.5 $&$  11.1 $&$  10.2 $&$  20.9 $&$  0.8 $&$  1.8 $&$  7.7 $&$  1.3 $&$  0.4 $&$  0.0 $&$  0.9 $&$  1.0 $&$  0.0 $&$  4.9 $&$  0.0 $&$  3.5 $\\
$0.0005 $&$ 11.5 $&$ 0.755-1.0 $&$ 0.0189 $&$   8.0 $&$  10.4 $&$  10.8 $&$  17.0 $&$  1.3 $&$  1.2 $&$  8.8 $&$  0.1 $&$  0.1 $&$  0.0 $&$  0.7 $&$  0.4 $&$  0.0 $&$  1.2 $&$  0.0 $ & $0.0$\\
$ 0.003 $&$  4.0 $&$ 0.033 $&$ 0.0159 $&$   6.6 $&$   4.4 $&$   4.3 $&$   8.3 $&$  1.8 $&$ -0.9 $&$ -0.5 $&$ -0.0 $&$ -3.6 $&$ -0.5 $&$ -0.9 $&$ -1.1 $&$  0.3 $&$ -0.3 $&$  0.0 $&$ -1.3 $\\
$ 0.003 $&$  4.0 $&$ 0.041 $&$ 0.0164 $&$   9.5 $&$   4.8 $&$   4.5 $&$  11.6 $&$  0.9 $&$ -0.6 $&$  0.5 $&$ -0.1 $&$ -2.6 $&$  0.0 $&$ -0.3 $&$ -1.9 $&$  0.0 $&$ -0.2 $&$  0.0 $&$ -0.7 $\\
$ 0.003 $&$  4.0 $&$ 0.054 $&$ 0.0160 $&$   7.4 $&$   3.7 $&$   4.7 $&$   9.5 $&$  1.0 $&$ -1.7 $&$ -0.6 $&$ -0.8 $&$ -1.1 $&$  0.0 $&$ -0.2 $&$ -2.5 $&$  0.0 $&$  0.3 $&$ -0.1 $&$  0.4 $\\
$ 0.003 $&$  4.0 $&$ 0.085 $&$ 0.0171 $&$  13.9 $&$   5.8 $&$   5.6 $&$  16.1 $&$  1.1 $&$ -2.3 $&$ -0.5 $&$ -0.2 $&$ -0.5 $&$  0.0 $&$ -0.2 $&$ -2.6 $&$  0.0 $&$  0.3 $&$  0.0 $&$  2.9 $\\
$ 0.003 $&$  4.0 $&$ 0.125 $&$ 0.0115 $&$  36.3 $&$  11.2 $&$   9.2 $&$  39.1 $&$ -3.0 $&$  5.3 $&$ -2.0 $&$  1.1 $&$ -0.2 $&$  0.0 $&$  0.0 $&$ -1.4 $&$  0.0 $&$  1.4 $&$  0.0 $&$  6.2 $\\
$ 0.003 $&$ 11.5 $&$ 0.089 $&$ 0.0222 $&$   9.2 $&$   3.4 $&$   3.2 $&$  10.3 $&$  0.8 $&$  0.8 $&$ -0.4 $&$  0.1 $&$ -0.7 $&$ -0.8 $&$ -0.9 $&$ -0.4 $&$  0.5 $&$ -0.5 $&$ -0.3 $&$  0.7 $\\
$ 0.003 $&$ 11.5 $&$ 0.101 $&$ 0.0227 $&$   7.5 $&$   3.2 $&$   2.9 $&$   8.7 $&$  0.8 $&$  1.4 $&$  0.1 $&$  0.2 $&$ -0.3 $&$ -0.2 $&$ -0.6 $&$ -0.1 $&$  0.1 $&$ -0.5 $&$ -0.1 $&$  0.7 $\\
$ 0.003 $&$ 11.5 $&$ 0.117 $&$ 0.0256 $&$   6.5 $&$   3.0 $&$   2.6 $&$   7.7 $&$  0.5 $&$  1.4 $&$  0.3 $&$  0.4 $&$ -0.2 $&$  0.0 $&$ -0.2 $&$  0.0 $&$  0.0 $&$ -0.4 $&$ -0.1 $&$  0.7 $\\
$ 0.003 $&$ 11.5 $&$ 0.155 $&$ 0.0288 $&$   3.4 $&$   2.4 $&$   2.8 $&$   5.0 $&$  1.1 $&$  1.1 $&$  0.1 $&$ -0.4 $&$ -0.1 $&$  0.0 $&$ -0.1 $&$ -0.4 $&$  0.0 $&$  0.1 $&$ -0.1 $&$  0.7 $\\
$ 0.003 $&$ 11.5 $&$ 0.244 $&$ 0.0281 $&$   3.5 $&$   2.4 $&$   2.6 $&$   4.9 $&$  0.7 $&$  1.1 $&$  0.1 $&$  0.1 $&$ -0.1 $&$  0.0 $&$ -0.1 $&$ -0.4 $&$  0.0 $&$  0.6 $&$ -0.1 $&$  0.5 $\\
$ 0.003 $&$ 11.5 $&$ 0.361 $&$ 0.0284 $&$   4.1 $&$   2.5 $&$   3.0 $&$   5.7 $&$ -1.0 $&$  1.3 $&$  0.5 $&$  0.7 $&$ -0.0 $&$  0.0 $&$ -0.0 $&$ -0.1 $&$  0.0 $&$ -0.1 $&$ -0.1 $&$  0.0 $\\
$ 0.003 $&$ 44.0 $&$ 0.341 $&$ 0.0379 $&$   8.6 $&$   3.6 $&$   2.6 $&$   9.7 $&$  1.1 $&$  0.9 $&$  0.2 $&$ -0.2 $&$  0.3 $&$ -0.7 $&$ -0.6 $&$  0.5 $&$  0.1 $&$ -0.0 $&$ -0.1 $&$  0.1 $\\
$ 0.003 $&$ 44.0 $&$ 0.386 $&$ 0.0350 $&$   8.0 $&$   3.4 $&$   3.1 $&$   9.2 $&$ -1.1 $&$  0.8 $&$  0.5 $&$ -0.0 $&$  0.1 $&$ -0.2 $&$ -0.1 $&$  0.2 $&$  0.0 $&$ -0.2 $&$ -0.2 $&$  1.2 $\\
$ 0.003 $&$ 44.0 $&$ 0.446 $&$ 0.0316 $&$   7.7 $&$   3.2 $&$   4.0 $&$   9.2 $&$  0.6 $&$  1.4 $&$  0.8 $&$  0.3 $&$  0.0 $&$  0.0 $&$ -0.1 $&$  0.2 $&$  0.0 $&$ -0.1 $&$ -0.5 $&$  2.6 $\\
$ 0.003 $&$ 44.0 $&$ 0.592 $&$ 0.0412 $&$   4.1 $&$   2.5 $&$   5.6 $&$   7.4 $&$  0.4 $&$  1.4 $&$  1.5 $&$ -0.2 $&$ -0.0 $&$  0.0 $&$  0.0 $&$ -0.1 $&$  0.0 $&$ -0.0 $&$ -1.2 $&$  4.8 $\\
$ 0.003 $&$ 44.0 $&$ 0.76-1.0 $&$ 0.0148 $&$   4.8 $&$   2.6 $&$   9.8 $&$  11.2 $&$ -0.3 $&$  1.7 $&$  2.1 $&$ -0.0 $&$ -0.0 $&$  0.0 $&$  0.0 $&$ -0.1 $&$  0.0 $&$ -0.3 $&$ -1.1 $& $0.0$\\
  \hline
\end{tabular}
\end{center} 
\end{footnotesize} 
\caption{
  The diffractive reduced cross section $\sigma_r^D$ at $\sqrt{s} = 252\gev$, 
  multiplied by $\xpom$, measured with the $575\gev$ data, at fixed values of 
  $\xpom$, $Q^2$ and $\beta$.  At the largest $\beta$, the bin-averaged cross section
  is given together with the lower and upper bin boundaries.  The statistical ($\delta_{stat}$), 
  uncorrelated ($\delta_{unc}$) and sum of all correlated ($\delta_{cor}$) uncertainties are given 
  together with the total uncertainty ($\delta_{tot}$).  The other columns show the individual 
  correlated uncertainties, which are due to the positron energy scale ($\delta_{ele}$), the positron
  polar angle measurement ($\delta_{\theta}$), the LAr noise subtraction ($\delta_{noi}$), the 
  hadronic SpaCal energy scale ($\delta_{spa}$), the efficiency of the logarithmic energy-weighted
  cluster radius cut ($\delta_{r_{log}}$), the charge asymmetry of the photoproduction background 
  ($\delta_{asy}$), the model uncertainty due to the influence of $F_L^D$ ($\delta_{mod}$), the model
  uncertainty on the underlying $\beta$ and $\xpom$ distributions ($\delta_{\beta}$, $\delta_{\xpom}$), 
  the influence of resonant ($\delta_{vm}$) and QED Compton ($\delta_{Com}$) contributions and finally
  the parametrisation choice for the bin centre corrections ($\delta_{bcc}$). A minus sign indicates
  that a source is anti-correlated with a change in the cross section. All uncertainties are 
given in per cent. The normalisation uncertainty of $8.1\%$ is not included.
}
\label{tab:3}
\end{table} 

\begin{table} 
\begin{footnotesize} 
\begin{center}
\begin{tabular}{|c | c | c || c | c | c | c | c || c | c | c | c | c | c | c | c | c | c | c | c |} 
  \hline 
  $\xpom$ & $Q^2$ & $\beta$ & $\xpom\sigma_r^D$ & $\delta_{stat}$ & $\delta_{unc}$ & $\delta_{cor}$ & $\delta_{tot}$ & $\delta_{ele}$ & $\delta_{\theta}$ & $\delta_{noi}$ & $\delta_{spa}$ & $\delta_{r_{log}}$ & $\delta_{asy}$ & $\delta_{mod}$ & $\delta_{\beta}$ & $\delta_{\xpom}$ & $\delta_{vm}$ & $\delta_{com}$ & $\delta_{bcc}$ \\
  & $[\gevsq]$ & & & $[\%]$ & $[\%]$ & $[\%]$ & $[\%]$ & $[\%]$ & $[\%]$ & $[\%]$ & $[\%]$ & $[\%]$ & $[\%]$ & $[\%]$ & $[\%]$ & $[\%]$ & $[\%]$ & $[\%]$ & $[\%]$\\ \hline\hline
$0.0005 $&$ 11.5 $&$ 0.570 $&$ 0.0553 $&$   1.3 $&$  10.4 $&$   6.9 $&$   8.7 $&$  0.4 $&$  2.7 $&$  4.0 $&$  1.1 $&$  0.2 $&$  0.2 $&$  0.2 $&$  0.8 $&$  0.0 $&$  3.8 $&$ -0.1 $&$  1.1 $\\
$0.0005 $&$ 11.5 $&$ 0.699 $&$ 0.0579 $&$   1.6 $&$  10.1 $&$   6.9 $&$  12.3 $&$  0.6 $&$  3.1 $&$  3.9 $&$  0.7 $&$  0.1 $&$  0.1 $&$  0.1 $&$  0.4 $&$  0.0 $&$  2.8 $&$ -0.2 $&$  3.2 $\\
$0.0005 $&$ 11.5 $&$ 0.755-1.0 $&$ 0.0198 $&$   1.2 $&$  10.0 $&$   8.5 $&$  13.2 $&$ -0.4 $&$  3.4 $&$  4.6 $&$  0.5 $&$  0.0 $&$  0.0 $&$  0.2 $&$  0.2 $&$  0.0 $&$  0.8 $&$ -2.1 $& $0.0$\\
$ 0.003 $&$ 11.5 $&$ 0.089 $&$ 0.0271 $&$   1.4 $&$   2.2 $&$   3.0 $&$   4.0 $&$  0.3 $&$  2.2 $&$  0.1 $&$  0.2 $&$  0.0 $&$  0.0 $&$ -0.2 $&$ -0.1 $&$  0.0 $&$ -0.3 $&$ -0.2 $&$  0.5 $\\
$ 0.003 $&$ 11.5 $&$ 0.101 $&$ 0.0275 $&$   1.3 $&$   2.2 $&$   3.1 $&$   4.0 $&$  0.1 $&$  2.2 $&$  0.1 $&$  0.1 $&$  0.1 $&$  0.1 $&$ -0.4 $&$ -0.2 $&$  0.0 $&$ -0.2 $&$ -0.2 $&$  0.6 $\\
$ 0.003 $&$ 11.5 $&$ 0.117 $&$ 0.0268 $&$   1.2 $&$   2.2 $&$   3.2 $&$   4.0 $&$ -0.3 $&$  2.3 $&$ -0.2 $&$  0.1 $&$  0.0 $&$  0.0 $&$ -0.3 $&$ -0.2 $&$  0.0 $&$ -0.1 $&$ -0.1 $&$  0.6 $\\
$ 0.003 $&$ 11.5 $&$ 0.155 $&$ 0.0267 $&$   0.7 $&$   2.1 $&$   3.3 $&$   4.0 $&$ -0.3 $&$  2.4 $&$ -0.7 $&$  0.1 $&$  0.1 $&$  0.1 $&$ -0.2 $&$ -0.2 $&$  0.0 $&$  0.3 $&$ -0.1 $&$  0.7 $\\
$ 0.003 $&$ 11.5 $&$ 0.244 $&$ 0.0270 $&$   0.7 $&$   2.1 $&$   3.4 $&$   4.0 $&$  0.3 $&$  2.5 $&$ -0.8 $&$  0.2 $&$  0.2 $&$  0.2 $&$  0.2 $&$  0.2 $&$  0.0 $&$  0.5 $&$ -0.3 $&$  0.6 $\\
$ 0.003 $&$ 11.5 $&$ 0.361 $&$ 0.0313 $&$   1.3 $&$   2.2 $&$   3.8 $&$   4.6 $&$  2.5 $&$  2.0 $&$  0.4 $&$  0.3 $&$  0.3 $&$  0.3 $&$  0.3 $&$  0.3 $&$  0.0 $&$ -0.3 $&$ -0.2 $&$ -0.5 $\\
$ 0.003 $&$ 44.0 $&$ 0.341 $&$ 0.0377 $&$   1.8 $&$   2.3 $&$   2.9 $&$   4.1 $&$  0.4 $&$  1.3 $&$  0.6 $&$  0.4 $&$  0.4 $&$  0.4 $&$  0.4 $&$  0.4 $&$  0.0 $&$ -0.0 $&$ -1.4 $&$  0.8 $\\
$ 0.003 $&$ 44.0 $&$ 0.386 $&$ 0.0389 $&$   1.7 $&$   2.3 $&$   3.6 $&$   4.6 $&$ -0.2 $&$  1.4 $&$  0.7 $&$  0.1 $&$  0.1 $&$  0.1 $&$  0.1 $&$  0.1 $&$  0.0 $&$ -0.3 $&$ -1.9 $&$  1.6 $\\
$ 0.003 $&$ 44.0 $&$ 0.446 $&$ 0.0410 $&$   1.4 $&$   2.2 $&$   4.0 $&$   4.8 $&$ -0.5 $&$  1.5 $&$  0.7 $&$  0.1 $&$  0.1 $&$  0.1 $&$  0.1 $&$  0.1 $&$  0.0 $&$  0.1 $&$ -1.3 $&$  2.8 $\\
$ 0.003 $&$ 44.0 $&$ 0.592 $&$ 0.0404 $&$   0.9 $&$   2.1 $&$   5.5 $&$   6.0 $&$ -0.5 $&$  1.6 $&$  0.8 $&$  0.1 $&$  0.1 $&$  0.1 $&$ -0.1 $&$  0.1 $&$  0.0 $&$ -0.0 $&$ -1.3 $&$  4.8 $\\
$ 0.003 $&$ 44.0 $&$ 0.76-1.0 $&$ 0.0162 $&$   1.0 $&$   2.1 $&$   9.7 $&$  10.0 $&$ -0.3 $&$  1.7 $&$  1.6 $&$ -0.5 $&$  0.1 $&$  0.1 $&$  0.1 $&$  0.1 $&$  0.0 $&$ -0.3 $&$ -0.9 $& $0.0$\\
  \hline
\end{tabular}
\end{center} 
\end{footnotesize} 
\caption{
  The diffractive reduced cross section $\sigma_r^D$ at $\sqrt{s} = 319\gev$, 
  multiplied by $\xpom$, measured with the $920\gev$ data, at fixed values of 
  $\xpom$, $Q^2$ and $\beta$.  At the largest $\beta$, the bin-averaged cross section
  is given together with the lower and upper bin boundaries.  The statistical ($\delta_{stat}$), 
  uncorrelated ($\delta_{unc}$) and sum of all correlated ($\delta_{cor}$)
uncertainties are given 
  together with the total uncertainty ($\delta_{tot}$).  The other columns show the individual 
  correlated uncertainties, which are due to the positron energy scale ($\delta_{ele}$), the positron
  polar angle measurement ($\delta_{\theta}$), the LAr noise subtraction ($\delta_{noi}$), the 
  hadronic SpaCal energy scale ($\delta_{spa}$), the efficiency of the logarithmic energy-weighted
  cluster radius cut ($\delta_{r_{log}}$), the charge asymmetry of the photoproduction background 
  ($\delta_{asy}$), the model uncertainty due to the influence of $F_L^D$ ($\delta_{mod}$), the model
  uncertainty on the underlying $\beta$ and $\xpom$ distributions ($\delta_{\beta}$, $\delta_{\xpom}$), 
  the influence of resonant ($\delta_{vm}$) and QED Compton ($\delta_{Com}$) contributions and finally
  the parametrisation choice for the bin centre corrections ($\delta_{bcc}$). A minus sign indicates
  that a source is anti-correlated with a change in the cross section. All uncertainties are 
given in per cent. The normalisation uncertainty of $7.6\%$ is not included.
}
\label{tab:4}
\end{table}

\begin{table} 
\begin{footnotesize} 
\begin{center}
\begin{tabular}{|c | c | c|| c | c | c | c | c || c | c | c | c | c |} 
\hline
$\xpom$ & $Q^2$ & $\beta$ & $\xpom F_L^D$ & $\delta_{stat}$ & $\delta_{stat+unc}$ & $\delta_{cor}$ & $\delta_{tot}$ & $\xpom F_2^D$ & $\delta_{stat}$ & $\delta_{stat+unc}$ & $\delta_{cor}$ & $\delta_{tot}$ \\
 & $[\gevsq]$ & & && &  &  & & &  &  &  \\ \hline
\hline 
$0.0005 $&$  4.0 $&$ 0.227 $&$ 0.0344 $&$ 0.0089 $&$ 0.0122 $&$ 0.0070 $&$ 0.0141 $&$ 0.0331 $&$ 0.0025 $&$ 0.0038 $&$ 0.0004 $&$ 0.0038 $\\
$0.0005 $&$ 11.5 $&$ 0.570 $&$ 0.0219 $&$ 0.0103 $&$ 0.0146 $&$ 0.0083 $&$ 0.0168 $&$ 0.0557 $&$ 0.0015 $&$ 0.0044 $&$ 0.0028 $&$ 0.0053 $\\
$0.0005 $&$ 11.5 $&$ 0.699 $&$ -0.0118 $&$ 0.0249 $&$ 0.0382 $&$ 0.0237 $&$ 0.0449 $&$ 0.0527 $&$ 0.0021 $&$ 0.0063 $&$ 0.0015 $&$ 0.0065 $\\
$ 0.003 $&$  4.0 $&$ 0.033 $&$ 0.0152 $&$ 0.0038 $&$ 0.0044 $&$ 0.0018 $&$ 0.0048 $&$ 0.0211 $&$ 0.0017 $&$ 0.0020 $&$ 0.0004 $&$ 0.0020 $\\
$ 0.003 $&$  4.0 $&$ 0.041 $&$ 0.0202 $&$ 0.0055 $&$ 0.0065 $&$ 0.0021 $&$ 0.0069 $&$ 0.0205 $&$ 0.0015 $&$ 0.0018 $&$ 0.0002 $&$ 0.0018 $\\
$ 0.003 $&$  4.0 $&$ 0.054 $&$ 0.0309 $&$ 0.0086 $&$ 0.0103 $&$ 0.0029 $&$ 0.0107 $&$ 0.0190 $&$ 0.0013 $&$ 0.0015 $&$ 0.0001 $&$ 0.0015 $\\
$ 0.003 $&$ 11.5 $&$ 0.089 $&$ 0.0103 $&$ 0.0039 $&$ 0.0043 $&$ 0.0022 $&$ 0.0048 $&$ 0.0275 $&$ 0.0007 $&$ 0.0010 $&$ 0.0007 $&$ 0.0013 $\\
$ 0.003 $&$ 11.5 $&$ 0.101 $&$ 0.0191 $&$ 0.0034 $&$ 0.0041 $&$ 0.0016 $&$ 0.0044 $&$ 0.0285 $&$ 0.0006 $&$ 0.0009 $&$ 0.0008 $&$ 0.0012 $\\
$ 0.003 $&$ 11.5 $&$ 0.117 $&$ 0.0105 $&$ 0.0044 $&$ 0.0055 $&$ 0.0016 $&$ 0.0057 $&$ 0.0267 $&$ 0.0005 $&$ 0.0009 $&$ 0.0007 $&$ 0.0011 $\\
$ 0.003 $&$ 11.5 $&$ 0.155 $&$ 0.0054 $&$ 0.0050 $&$ 0.0077 $&$ 0.0039 $&$ 0.0086 $&$ 0.0263 $&$ 0.0003 $&$ 0.0007 $&$ 0.0008 $&$ 0.0011 $\\
$ 0.003 $&$ 44.0 $&$ 0.341 $&$ 0.0163 $&$ 0.0078 $&$ 0.0085 $&$ 0.0026 $&$ 0.0089 $&$ 0.0388 $&$ 0.0013 $&$ 0.0018 $&$ 0.0012 $&$ 0.0021 $\\
$ 0.003 $&$ 44.0 $&$ 0.386 $&$ 0.0086 $&$ 0.0064 $&$ 0.0075 $&$ 0.0027 $&$ 0.0080 $&$ 0.0384 $&$ 0.0010 $&$ 0.0015 $&$ 0.0014 $&$ 0.0020 $\\
$ 0.003 $&$ 44.0 $&$ 0.446 $&$ 0.0298 $&$ 0.0070 $&$ 0.0086 $&$ 0.0033 $&$ 0.0092 $&$ 0.0414 $&$ 0.0009 $&$ 0.0014 $&$ 0.0015 $&$ 0.0021 $\\
$ 0.003 $&$ 44.0 $&$ 0.592 $&$ 0.0066 $&$ 0.0090 $&$ 0.0129 $&$ 0.0039 $&$ 0.0134 $&$ 0.0395 $&$ 0.0005 $&$ 0.0012 $&$ 0.0021 $&$ 0.0024 $\\

\hline 
\end{tabular}
\end{center} 
\end{footnotesize} 
\caption{The diffractive structure functions $F_L^D$ and $F_2^D$ multiplied by $\xpom$, at fixed values of $\xpom$, $Q^2$ and $\beta$.
  The statistical uncertainty ($\delta_{stat}$), the sum of the statistical and uncorrelated uncertainties ($\delta_{stat+unc}$) and the sum of all correlated uncertainties ($\delta_{cor}$)
  are given together with the total uncertainty ($\delta_{tot}$). Absolute uncertainties are given.
  The normalisation uncertainty of $8.1\%$ is not included.}
\label{tab:5}
\end{table} 

\begin{table} 
\begin{footnotesize} 
\begin{center}
\begin{tabular}{|c | c | c|| c | c | c | c | c |} 
\hline 
$\xpom$ & $Q^2$ & $\beta$ & $R^D$ & $\delta_{stat}$ & $\delta_{stat+unc}$ & $\delta_{cor}$ & $\delta_{tot}$ \\
 & $[\gevsq]$ & & & & & & \\ \hline \hline
$0.0005 $&$  4.0 $&$ 0.227 $&$ 258.3940 $&$ 331.9017 $&$ 666.2490 $&$ 495.2585 $&$ 830.1619 $\\
$0.0005 $&$ 11.5 $&$ 0.570 $&$ 0.6477 $&$ 0.5000 $&$ 0.6475 $&$ 0.3902 $&$ 0.7560 $\\
$0.0005 $&$ 11.5 $&$ 0.699 $&$ -0.1829 $&$ 0.4141 $&$ 0.5014 $&$ 0.2941 $&$ 0.5813 $\\
$ 0.003 $&$  4.0 $&$ 0.033 $&$ 2.5995 $&$ 1.6409 $&$ 1.9328 $&$ 1.1427 $&$ 2.2454$\\
$ 0.003 $&$  4.0 $&$ 0.041 $&$ 65.2663 $&$ 501.4577 $&$ 783.3735 $&$ 168.2953 $&$ 801.2474 $\\
$ 0.003 $&$  4.0 $&$ 0.054 $&$ 207.3901 $&$ 191.8850 $&$ 241.6402 $&$ 100.8154 $&$ 261.8276 $\\
$ 0.003 $&$ 11.5 $&$ 0.089 $&$ 0.5942 $&$ 0.3415 $&$ 0.3723 $&$ 0.1871 $&$ 0.4166$\\
$ 0.003 $&$ 11.5 $&$ 0.101 $&$ 2.0212 $&$ 1.0084 $&$ 1.1574 $&$ 0.4189 $&$ 1.2308 $\\
$ 0.003 $&$ 11.5 $&$ 0.117 $&$ 0.6521 $&$ 0.4446 $&$ 0.5409 $&$ 0.1411 $&$ 0.5590$\\
$ 0.003 $&$ 11.5 $&$ 0.155 $&$ 0.2616 $&$ 0.2642 $&$ 0.4650 $&$ 0.2318 $&$ 0.5196$\\
$ 0.003 $&$ 44.0 $&$ 0.341 $&$ 0.7214 $&$ 0.5061 $&$ 0.6093 $&$ 0.1738 $&$ 0.6336$\\
$ 0.003 $&$ 44.0 $&$ 0.386 $&$ 0.2905 $&$ 0.2702 $&$ 0.3150 $&$ 0.1084 $&$ 0.3331$\\
$ 0.003 $&$ 44.0 $&$ 0.446 $&$ 2.5776 $&$ 2.3446 $&$ 2.4918 $&$ 0.8133 $&$ 2.6211 $\\
$ 0.003 $&$ 44.0 $&$ 0.592 $&$ 0.2014 $&$ 0.3133 $&$ 0.4689 $&$ 0.1415 $&$ 0.4898$\\
\hline 
\end{tabular}
\end{center} 
\end{footnotesize} 
\caption{The ratio $R_D$ of the cross sections for longitudinally to transversely polarised photon 
cross sections, at fixed values of $\xpom$, $Q^2$ and $\beta$.
  The sum of the statistical and uncorrelated uncertainties ($\delta_{stat+unc}$) and the sum of all correlated uncertainties ($\delta_{cor}$)
  are given together with the total uncertainty ($\delta_{tot}$). Absolute uncertainties are give.
}
\label{tab:6}
\end{table} 
\end{landscape}

\pagebreak

\begin{thebibliography}{99}

\bibitem{ref1} 
  M.~Derrick {\it et al.}  [ZEUS Collaboration],
  Phys.\ Lett.\ B {\bf 315} (1993) 481; \\
  T.~Ahmed {\it et al.}  [H1 Collaboration],
  Nucl.\ Phys.\ B {\bf 429} (1994) 477.

\bibitem{ref5}
  J.~Collins,
  Phys.\ Rev.\ D {\bf 57} (1998) 3051
  [Erratum-ibid.\ D {\bf 61} (2000) 019902]
  [hep-ph/9709499].

\bibitem{ref6} 
  A.~Aktas {\it et al.}  [H1 Collaboration],
  Eur.\ Phys.\ J.\  C {\bf 48} (2006) 715
  [hep-ex/0606004].

\bibitem{ref7} 
  A.~Aktas {\it et al.}  [H1 Collaboration],
  Eur.\ Phys.\ J.\  C {\bf 48} (2006) 749
  [hep-ex/0606003].

\bibitem{ref8}
  F.~Aaron {\it et al.}  [H1 Collaboration],
  Eur.\ Phys.\ J.\  C {\bf 71} (2011) 1578
  [arXiv:1010.1476].


\bibitem{ref9} 
  S.~Chekanov {\it et al.}  [ZEUS Collaboration],
  Nucl.\ Phys.\  B {\bf 816} (2009) 1 [arXiv:0812.2003].

\bibitem{ref10}
  S.~Chekanov {\it et al.} [ZEUS Collaboration],
  Nucl.\ Phys.\  B {\bf 831} (2010) 1
  [arXiv:0911.4119].

\bibitem{ref11} 
  A.~Martin, M.~Ryskin and G.~Watt,
  Eur.\ Phys.\ J.\ C {\bf 44} (2005) 69
  [hep-ph/0504132].

\bibitem{ref12} 
  A.~Aktas {\it et al.}  [H1 Collaboration],
  JHEP {\bf 0710} (2007) 042
  [arXiv:0708.3217].

\bibitem{ref13} 
A.~Aktas {\it et al.}  [H1 Collaboration],
Eur.\ Phys.\ J.\  C {\bf 50} (2007) 1 [hep-ex/0610076]; \\
  S.~Chekanov {\it et al.}  [ZEUS Collaboration],
  Nucl.\ Phys.\  B {\bf 672} (2003) 3 [hep-ex/0307068].


\bibitem{ref14}
  J.~Bl\"umlein and D.~Robaschik,
  Phys.\ Lett.\  B {\bf 517} (2001) 222
  [hep-ph/0106037]; \\
  J.~Bl\"umlein, B.~Geyer and D.~Robaschik,
  Nucl.\ Phys.\  B {\bf 755} (2006) 112
  [hep-ph/0605310]; \\
  J.~Bl\"umlein, D.~Robaschik and B.~Geyer,
  Eur.\ Phys.\ J.\  C {\bf 61} (2009) 279
  [arXiv:0812.1899].

\bibitem{ref15}
  A.~Zee, F.~Wilczek and S.~B.~Treiman,
  Phys.\ Rev.\  D {\bf 10} (1974) 2881; \\
  G.~Altarelli and G.~Martinelli,
  Phys.\ Lett.\  B {\bf 76} (1978) 89.


\bibitem{ref2}
  L.~V.~Gribov, E.~M.~Levin and M.~G.~Ryskin,
  Phys.\ Rept.\  {\bf 100} (1983) 1; \\
  A.~H.~Mueller,
  Nucl.\ Phys.\  B {\bf 335} (1990) 115; \\
  N.~N.~Nikolaev and B.~G.~Zakharov,
  Z.\ Phys.\  C {\bf 49} (1991) 607.

\bibitem{ref16} 
  E.~A.~Kuraev, L.~N.~Lipatov and V.~S.~Fadin,
  Sov.\ Phys.\ JETP {\bf 44} (1976) 443
  [Zh.\ Eksp.\ Teor.\ Fiz.\  {\bf 71} (1976) 840]; \\
  E.~A.~Kuraev, L.~N.~Lipatov and V.~S.~Fadin,
  Sov.\ Phys.\ JETP {\bf 45} (1977) 199
  [Zh.\ Eksp.\ Teor.\ Fiz.\  {\bf 72} (1977) 377]; \\
  I.~I.~Balitsky and L.~N.~Lipatov,
  Sov.\ J.\ Nucl.\ Phys.\  {\bf 28} (1978) 822
  [Yad.\ Fiz.\  {\bf 28} (1978) 1597].

\bibitem{ref17}
  M.~Ciafaloni,
  Nucl.\ Phys.\ B {\bf 296} (1988) 49; \\
  S.~Catani, F.~Fiorani and G.~Marchesini,
  Phys.\ Lett.\ B {\bf 234} (1990) 339; \\
  S.~Catani, F.~Fiorani and G.~Marchesini,
  Nucl.\ Phys.\ B {\bf 336} (1990) 18; \\
  G.~Marchesini,
  Nucl.\ Phys.\ B {\bf 445} (1995) 49
  [hep-ph/9412327].

\bibitem{ref18}
  S.~Chekanov {\it et al.}  [ZEUS Collaboration],
  Eur.\ Phys.\ J.\  C {\bf 38} (2004) 43
  [hep-ex/0408009].

\bibitem{ref19} 
  M.~Diehl, Proceedings of the Blois Workshop on
  Elastic and Diffractive Scattering, Blois, France, May 2005,
  hep-ph/0509107.

\bibitem{ref20} 
  F.~D.~Aaron {\it et al.}  [H1 Collaboration],
  Phys.\ Lett.\  B {\bf 665} (2008) 139
  [arXiv:0805.2809].

\bibitem{ref21}
  F.~D.~Aaron {\it et al.}  [H1 Collaboration],
  Eur.\ Phys.\ J.\  C {\bf 71} (2011) 1579
  [arXiv:1012.4355].

\bibitem{ref22} 
  P.~R.~Newman,
  Proceedings of the `HERA and the LHC' Workshop, 
eds. A.~de Roeck, H.~Jung,  
  CERN-2005-14 (2005) 514 
  [hep-ex/0511047].

\bibitem{ref23}
  F.~D.~Aaron {\it et al.}  [H1 Collaboration],
  Eur.\ Phys.\ J.\  C {\bf 63} (2009) 625
  [arXiv:0904.0929].

\bibitem{ref24} 
  F.~D.~Aaron {\it et al.}  [H1 Collaboration],
  JHEP {\bf 1005} (2010) 032
  [arXiv:0910.5831].

\bibitem{ref25}
  R.~Brock {\it et al.}  [CTEQ Collaboration],
  Rev.\ Mod.\ Phys.\  {\bf 67} (1995) 157.


\bibitem{ref4}
  K.~J.~Golec-Biernat and M.~W\"usthoff,
  Phys.\ Rev.\  D {\bf 59} (1998) 014017
  [hep-ph/9807513]; \\
  K.~J.~Golec-Biernat and M.~W\"usthoff,
  Phys.\ Rev.\  D {\bf 60} (1999) 114023
  [hep-ph/9903358].

\bibitem{ref3}
  H.~Kowalski, L.~Motyka and G.~Watt,
  Phys.\ Rev.\  D {\bf 74} (2006) 074016
  [hep-ph/0606272].

\bibitem{ref26}
  J.~Bartels, J.~R.~Ellis, H.~Kowalski and M.~W\"usthoff,
  Eur.\ Phys.\ J.\  C {\bf 7} (1999) 443
  [hep-ph/9803497].

\bibitem{ref27}
  A.~Hebecker and T.~Teubner,
  Phys.\ Lett.\  B {\bf 498} (2001) 16
  [hep-ph/0010273].

\bibitem{ref28}
  M.~G.~Ryskin,
  Z.\ Phys.\  C {\bf 57} (1993) 89; \\
  A.~D.~Martin, M.~G.~Ryskin and T.~Teubner,
  Phys.\ Rev.\  D {\bf 62} (2000) 014022
  [hep-ph/9912551].

\bibitem{ref29}
  K.~J.~Golec-Biernat and A.~\L uszczak,
  Phys.\ Rev.\  D {\bf 76} (2007) 114014
  [arXiv:0704.1608].

\bibitem{ref30}
I.~Abt {\it et al.}  [H1 Collaboration],
  Nucl.\ Instrum.\ Meth.\  A {\bf 386} (1997) 310; \\
  I.~Abt {\it et al.}  [H1 Collaboration],
  Nucl.\ Instrum.\ Meth.\  A {\bf 386} (1997) 348; \\
  R.~Appuhn {\it et al.}  [H1 SPACAL Group],
  Nucl.\ Instrum.\ Meth.\ A {\bf 386} (1997) 397.

\bibitem{cst}
D. Pitzl {\it et~al.}, {Nucl.\ Instr.\ and Meth.\ A}{\bf 454} (2000) 334 
  [hep-ex/0002044].

\bibitem{Becker:2007ms}
J. Becker {\it et~al.}, {Nucl.\ Instrum.\ Meth.\ A} {\bf 586} (2008) 190 
 [arXiv:physics/0701002].

\bibitem{ref31}
  B.~Andrieu {\it et al.}  [H1 Calorimeter Group],
  Nucl.\ Instrum.\ Meth.\ A {\bf 350} (1994) 57.\\
  B.~Andrieu {\it et al.}  [H1 Calorimeter Group],
  Nucl.\ Instrum.\ Meth.\  A {\bf 336} (1993) 499.

\bibitem{ref32} 
  S. Piec, Doctoral thesis, Humboldt-Universit\"{a}t zu Berlin (2009) 
  "Measurement of the Proton Structure Function $F_L(x,Q^2)$ with the H1
  Detector at HERA",
  [http://www-h1.desy.de/psfiles/theses/h1th-546.pdf].

\bibitem{ref33}
  M. Peez, Ph.D. thesis (in French), University of Lyon (2003),
  "Search for Deviations from the Standard Model in High Transverse
  Energy Processes at the Electron-Proton Collider HERA",
  [http://www-h1.desy.de/psfiles/theses/h1th-317.ps].

\bibitem{ref34} 
  C.~Adloff {\it et al.}  [H1 Collaboration],
  Z.\ Phys.\ C {\bf 76} (1997) 613
  [hep-ex/9708016].

\bibitem{ref35} 
  RAPGAP 3.1: H.~Jung,
  Comput.\ Phys.\ Commun.\  {\bf 86} (1995) 147.

\bibitem{ref37}
  M.~Bengtsson and T.~Sj\"{o}strand,
  Z.\ Phys.\ C {\bf 37} (1988) 465.

\bibitem{ref38}
  B.~Andersson, G.~Gustafson, G.~Ingelman and T.~Sj\"{o}strand,
  Phys.\ Rept.\  {\bf 97} (1983) 31.

\bibitem{ref39}
  T.~Sj\"{o}strand,
  Comput.\ Phys.\ Commun.\  {\bf 82} (1994) 74.

\bibitem{ref36} 
  B. List, A. Mastroberardino, 
  in A. Doyle {\it et al.} (eds.), Proceedings of the Workshop on
  Monte Carlo Generators for HERA Physics, DESY-PROC-1999-02 (1999)
  396.

\bibitem{ref40} 
G.~Sch\"{u}ler and H.~Spiesberger, Proc. of the Workshop on
Physics at HERA, eds. W.~Buchm\"uller, G.~Ingelman, Hamburg,
DESY (1992) 1419.

\bibitem{ref41}
  A.~Courau and P.~Kessler,
  Phys.\ Rev.\ D {\bf 46} (1992) 117.

\bibitem{ref42}
  R.~Brun {\it et al.}, CERN-DD/EE-84-1 (1987).


\bibitem{ref43} 
  D. Salek, Ph.D. thesis, Charles University, Prague (2010),
  "Measurement of the Longitudinal Proton Structure Function in
  Diffraction at the H1 Experiment and Prospects for Diffraction at
  LHC", [http://www-h1.desy.de/psfiles/theses/h1th-617.pdf].

\bibitem{ref44} 
  F.~D.~Aaron {\it et al.}  [H1 Collaboration],
  Eur.\ Phys.\ J.\  C {\bf 64} (2009) 561
  [arXiv:0904.3513].



\end{thebibliography}
\end{document}